\documentclass[man, a4paper,floatsintext]{apa6} 
\usepackage[american]{babel}
\usepackage[utf8]{inputenc}
\usepackage{graphicx}
\usepackage{csquotes}
\usepackage{amsmath}
\usepackage{amsthm}
\usepackage{amssymb}
\usepackage{bm,array}
\usepackage{latexsym}
\usepackage{setspace}
\usepackage{color}    
\usepackage{blkarray}
\usepackage{float}
\usepackage{forest}
\usetikzlibrary{arrows.meta, shapes.geometric, calc, shadows}

\definecolor{dkgreen}{rgb}{0,0.6,0}
\definecolor{darkblue}{rgb}{0,0,.4}
\definecolor{darkblack}{rgb}{0,0,.4}
\usepackage{listings}
\usepackage{setspace}
\AtBeginEnvironment{tabular}{\doublespacing}
\usepackage{microtype}
\usepackage{listings}             

\usepackage{natbib}
\DeclareFontFamily{U}{mathx}{\hyphenchar\font45}
\DeclareFontShape{U}{mathx}{m}{n}{
      <5> <6> <7> <8> <9> <10>
      <10.95> <12> <14.4> <17.28> <20.74> <24.88>
      mathx10
      }{}
\DeclareSymbolFont{mathx}{U}{mathx}{m}{n}
\DeclareMathSymbol{\bigtimes}{1}{mathx}{"91}

\definecolor{pigment}{rgb}{0.2, 0.2, 0.6}
\usepackage{hyperref}
\hypersetup{
  colorlinks   = true, 
  urlcolor     = pigment, 
  linkcolor    = pigment, 
  citecolor   = pigment 
}
\newcommand*{\doi}[1]{\href{http://dx.doi.org/\detokenize{#1} {\raggedright\mybibdoicolor{DOI: \detokenize{#1}}}}}

\makeatletter
\let\amsmath@bigm\bigm

\newcommand*\diff{\mathop{}\!\mathrm{d}}

\renewcommand{\bigm}[1]{%
  \ifcsname fenced@\string#1\endcsname
    \expandafter\@firstoftwo
  \else
    \expandafter\@secondoftwo
  \fi
  {\expandafter\amsmath@bigm\csname fenced@\string#1\endcsname}%
  {\amsmath@bigm#1}%
}
\newcommand{\DeclareFence}[2]{\@namedef{fenced@\string#1}{#2}}

\makeatother
\DeclareFence{\mid}{|}

\title{
Multinomial Models with Linear Inequality Constraints:
Overview and Improvements of Computational Methods for Bayesian Inference
}

\shorttitle{Inequality-Constrained Multinomial Models}

\twoauthors{Daniel W. Heck}{Clintin P. Davis-Stober}
\twoaffiliations{University of Mannheim}{University of Missouri\vspace{-.3cm}}
\leftheader{Heck \& Davis-Stober}

\authornote{
The first author was supported by the research training group Statistical Modeling in Psychology (GRK 2277), funded by the German Research Foundation (DFG). The second author was supported by NSF grant SES 14-59866 (PI: C. Davis-Stober) and NIH grant (K25AA024182, PI: C. Davis-Stober).

The R package \texttt{multinomineq} can be installed from \href{https://github.com/danheck/multinomineq/}{https://github.com/danheck/multinomineq/}.
Data and R code for the analyses are available at the Open Science Framework at \href{https://osf.io/xv9u3/}{https://osf.io/xv9u3/}.

Correspondence concerning this article should be addressed to Daniel W. Heck, Department of Psychology, School of Social Sciences, University of Mannheim, B6 30-32 Room 313, D-68131 Mannheim,  Germany. E-mail: \href{mailto:heck@uni-mannheim.de}{heck@uni-mannheim.de}; Phone: (+49) 621 181 1891.
}

\abstract{
Many psychological theories can be operationalized as linear inequality constraints on the parameters of multinomial distributions (e.g., discrete choice analysis).  These constraints can be described in two equivalent ways:  Either as the solution set to a system of linear inequalities or as the convex hull of a set of extremal points (vertices).  For both representations, we describe a general Gibbs sampler for drawing posterior samples in order to carry out Bayesian analyses.  We also summarize alternative sampling methods for estimating Bayes factors for these model representations using the encompassing Bayes factor method.  We introduce the R package \texttt{multinomineq}, which provides an easily-accessible interface to a computationally efficient implementation of these techniques.
}


\begin{document}
\maketitle
\setcounter{secnumdepth}{3}

\section{Introduction}
\label{s.intro}

Multinomial random variables form the backbone of discrete and categorical data analysis within psychology and the behavioral sciences. 
The key to any viable data analysis is the successful translation of an abstract theoretical hypothesis into a concrete, statistical model. 
As a simple example, consider the hypothesis that overconsumption of drugs (i.e., taking more tablets than prescribed) decreases with the number of daily doses \citep{paes1997impact}.
To assess the validity of this prediction, one could test the statistical hypothesis that overconsumption is \emph{identical} across all dosage regimes. 
If this hypothesis is rejected, one could carry out subsequent analyses to determine if the rates differ across the dosage conditions in a pairwise fashion. 
Yet, testing the ``straw-man'' model of all dosage conditions resulting in identical rates of overconsumption is not necessarily a faithful translation of the original hypothesis, rather, it is a means to an end, serving only as a pretext to carrying out tests on multiple pairs of dosage conditions.  

To make this example more concrete, suppose we have three dosage regimes of the drug (i.e., once, twice, and three times daily) in a between-subjects design \citep{paes1997impact}. 
We model the number of participants showing overconsumption in each condition as a binomial random variable and define the parameters $\theta_1$, $\theta_2$, and $\theta_3$ as the corresponding probabilities that an individual takes more tablets than prescribed.
While we could test whether the three $\theta_i$ parameters are equal across all conditions (i.e., $\theta_{1}=\theta_{2}=\theta_{3}$), this does not directly follow from our original hypothesis which only specified a monotonic relationship between overconsumption and dosage regimen. 
Testing the hypothesis of interest requires specifying an \emph{ordering} relationship imposed on the overconsumption rates for each of the three dosage conditions: 
\begin{equation}
\label{e.drug}
\theta_{1} \geq \theta_{2} \geq \theta_{3}.
\end{equation}
Paired with the binomial likelihood function, these order constraints represent a more faithful statistical analysis of the hypothesis being tested \citep[see also][for a full discussion]{hoijtink2011informative}.
Testing order constraints such as these, and linear inequality constraints more generally, requires a bit more effort than simpler tests of equality, but, as we show, can be carried out efficiently and are more interpretable.  

A key difficulty in analyzing inequality-constrained models and theories is that it can quickly become difficult to characterize the resulting restricted parameter space \citep[e.g.,][]{fishburn1992induced, davis-stober2012lexicographic}.
Our drug dosage example is quite simple---indeed, for Eq.~\eqref{e.drug}, there are only two non-redundant pairwise order constraints, namely, $\theta_{1} \geq \theta_{2}$ and $\theta_{2} \geq \theta_{3}$.
When combined with the inequality constraints that the probability of overconsumption must be between zero and one for all conditions (i.e., $0 \leq \theta_{i} \leq 1$), this completely characterizes the ordering relationships of interest.  
However, not all interesting hypotheses are so simple in structure.
As we illustrate in Section~\ref{s.ex_transitivity}, the random preference model of \citet{regenwetter2012behavioral} is far more complex with 75,834 non-redundant linear inequalities.

In general, bounded, linearly restricted parameter spaces can be defined in two different, yet equivalent, ways \citep[][]{brondsted2012introduction}. 
First, the restricted parameter space can be defined as the solution space to a system of a finite number of linear inequalities and equalities - similar to our drug dosage example.  
Alternatively, the same restricted parameter space can be defined as the convex hull of a set of extremal points (vertices).  
Let $\bm\theta = (\theta_{1},\theta_{2},\theta_{3})$. 
For our simple dosage example, the set of all extremal points is the set of all vectors, $\bm\theta$, where each entry is equal to 0 or 1 and satisfy the above inequalities, which yields the set: $(0,0,0)$, $(1,0,0)$, $(1,1,0)$, and $(1,1,1)$.
Section~\ref{s.theory} shows that it is often relatively easy to derive these vertices by enumerating all patterns that are predicted by a psychological theory even though it may be difficult to specify the corresponding system of inequality constraints \citep{regenwetter2017constructbehavior}.

Irrespective of how inequality constraints are formally specified, their statistical analysis has been a long-standing issue in mathematical psychology \citep{iverson1985statistical} and statistics in general \citep{silvapulle2004constrained, barlow1972statistical, robertson1988order}.
In classical statistics, basic results regarding the asymptotic distribution of the likelihood ratio test are valid when testing equality constraints, but are not when testing inequality constraints \citep{davis-stober2009analysis, silvapulle2004constrained}.
As a remedy, methods for inequality-constrained models have recently been developed in the Bayesian framework \citep{myung2005bayesian, karabatsos2005exchangeable, klugkist2005inequality, hoijtink2008bayesian, sedransk1985bayesian} or based on minimum description length  \citep{klauer2015flexibility, heck2015testing, rissanen1978modeling}.
Multinomial models with inequality constraints have also been applied to the Bayesian analysis of contingency tables \citep[e.g.,][]{lindley1964bayesian, klugkist2010bayesian, agresti2005bayesian, laudy2007bayesian}. 
However, general-purpose software packages for Bayesian statistics such as JAGS \citep{plummer2003jags} or Stan \citep{stan} are often not suited for the analysis of models with complex inequality constraints.
This is due to the fact that the boundary of the constrained parameter space is specified as a, typically complex, function of multiple parameters. 
As a result, the parameters are highly inter-dependent and often cannot be defined independently \citep[for a counterexample with simple constraints, see][]{heck2016adjusted}.

This article considers computational methods of carrying out Bayesian analyses on multinomial models with linear inequality constraints on the parameters.
However, we go further than analyzing simple ``toy'' models such as the dosage example above and consider models defined by arbitrarily complex linear constraints on multinomial parameters.
Analyzing this class of model is known to be computationally challenging, especially for highly complex linear constraints as those defined by random preference models \citep{smeulders2018testing} and the axioms of additive conjoint measurement \citep{karabatsos2018bayesian}. 
In the following, Section~\ref{s.theory} highlights the relevance of inequality-constrained multinomial models for testing psychological theories.
In Section~\ref{s.multinomial}, we introduce the notation, likelihood, and prior for multinomial models and the two types of representations for inequality constraints. 
Section~\ref{s.Ab} extends existing computational methods for binomial models with specific order constraints \citep[e.g.,][]{myung2005bayesian, karabatsos2005exchangeable} to multinomial models with arbitrary sets of linear inequalities.
More precisely, we develop a general Gibbs sampler for parameter estimation and offer improved computational methods for estimating the encompassing Bayes factor for carrying out Bayesian model selection.
Section~\ref{s.V} develops these methods for models that are specified by a set of predicted patterns using the vertex representation.
This is useful, as defining a restricted model may be straightforward for one type of representation but not the other, while switching between representations can be computationally infeasible \citep{avis1997how}.
In Section~\ref{s.examples}, we offer the R package \texttt{multinomineq} \citep{heck2019multinomineq} and show how to apply inequality-constrained multinomial models in practice using concrete examples.
Finally, Section~\ref{s.discussion} discusses the analysis of nested data, the choice of priors, and possible directions for future research.

\subsection{Where Do Inequality Constraints Come From?}
\label{s.theory}


Inequality constraints on multinomial parameters can arise in a number of ways.  Similar to our drug consumption example, they can arise ``organically'' by directly instantiating the hypothesis of interest. For this example, the inequalities are implied by the natural hypothesis that the response categories should be ordered by dosage regimen.  In this way, inequality constraints can provide a direct evaluation of the hypothesis of interest, in contrast to other, heuristic methods such as testing the equality of all three dosage condition parameters and then carrying out additional, post hoc analyses to determine directional differences. In later sections, we will consider other examples of linear inequality constraints that arise naturally from theoretic hypotheses that are more complex than simple order restrictions \citep{hilbig2014generalized}.

While not immediately obvious, linear inequality constraints can also arise when evaluating theories/models/axioms in which \emph{multiple} predictions are made.  Such theories are quite common, especially in the field of judgment and decision making. For example, consider the well-known transitivity of preference axiom \citep{regenwetter2011transitivity}.  Depending upon an individual's tastes, there are many ways for a decision maker to have transitive preferences over a set of choice alternatives.  Evaluating multiple predictions of a theory simultaneously within a multinomial framework opens up additional ways to operationalize this theory of interest.  As an example, we consider methods of stochastic specification for deterministic theories, although we note that the application of such methods (e.g., mixture methods) extends beyond the decision making domain \citep{davis-stober2016bayes}.

\subsubsection{Stochastic Specification}
\label{s.stochastic}

Many psychological theories predict \emph{deterministic} choice patterns across different contexts (e.g., different types of stimuli, items, conditions, measurement occasions, or pre-existing groups). 
For instance, a theory might provide a specific response pattern such as ``participants prefer Option A over B in each of five choice scenarios'' \citep{broder2003bayesian}. 
Often, however, theories predict more than one response pattern.
As illustrated in Section~\ref{s.ex_DEgap} for the description-experience gap in the domain of risky gambles \citep{hertwig2004decisions}, the hypothesis that participants assign more weight to small probabilities results in multiple predicted patterns.  
The complete set of predicted patterns can be obtained in different ways \citep{regenwetter2017constructbehavior}, for instance, by (a) translating a verbal theory into predicted patterns, (b) deriving algebraic implications of axioms or formal theories, and (c) brute force enumeration of all of the predictions made by the deterministic theory, typically under a set of theory-specific assumptions (e.g., theory parameter values).%
	\footnote{For the brute force enumeration approach, this is typically handled by simulating the deterministic theory while systematically varying its parameter values.  To ensure accurate enumeration, a large number of simulation replications should be used to ensure that all possible predictions are simulated.}
Irrespective of how the theoretical predictions are derived, observed choice frequencies are inherently noisy and exhibit a certain amount of variance both within and across persons or contexts.  
Hence, the question arises of how to define a stochastic model for empirical frequencies based on a set of deterministic predicted patterns \citep{heck2017information, regenwetter2012behavioral, regenwetter2018role, carbone2000which}.

\begin{figure}[th]
\centering
\includegraphics[width=\linewidth]{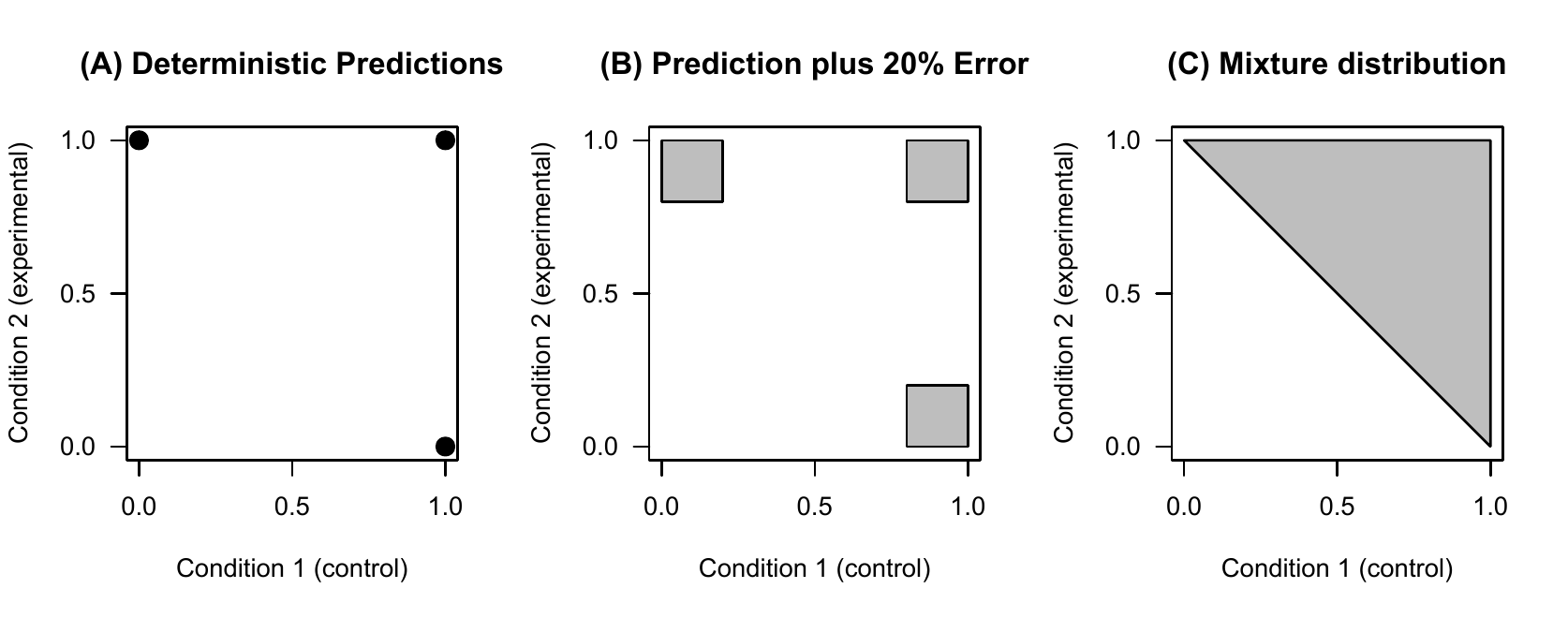}
\caption{
(A) Many psychological theories predict multiple deterministic response patterns (black points).
(B) To obtain a stochastic model for observed frequencies, some theories assume that one of the patterns underlies all observations, with independent error probabilities smaller than 20\% (gray boxes).
(C) Other theories assume random variation in the data-generating pattern across observations, which results in response probabilities that are defined by the mixture distribution over the three predicted patterns (gray triangle).
}
\label{f.pred}
\end{figure}

In multinomial models, each predicted choice pattern can be represented by a vector of probabilities of either one (an option is deterministically chosen) or zero \citep[an option is not chosen;][]{broder2003bayesian}.
Figure~\ref{f.pred} illustrates this for two independent binomial probabilities $\bm \theta = (\theta_{1}, \theta_{2})$ of preferring Option A over B in a control and an experimental condition, respectively.
The three black points in Figure~\ref{f.pred}A show three predicted patterns of a hypothetical theory that are represented by the vectors $\bm v^{(1)} = (0,1)$, $\bm v^{(2)} = (1,1)$, and $\bm v^{(3)} = (1,0)$.
For instance, the pattern $\bm v^{(3)} = (1,0)$ represents the prediction that Option A is chosen in the control condition (since $\theta_{1}=1$) whereas Option B is chosen in the experimental condition (since $\theta_{2}=0$).

To derive a stochastic model based on a set of predictions $\bm v^{(s)}$, it is important to consider \textit{why} a psychological theory makes multiple predictions in the first place \citep{regenwetter2017constructbehavior}.
A theory might assume that one of the predicted patterns consistently describes the ``true'' data-generating mechanism across all measurement occasions.
According to this interpretation, theory-inconsistent responses merely emerge from unsystematic errors in responding (e.g., due to inattention) whereas latent preferences are stable.
In our example, this assumption results in a stochastic model with two independent error probabilities for the two conditions.
These error probabilities serve as free parameters and are usually constrained to be below a predefined, fixed threshold such as 20\%.
In Figure~\ref{f.pred}B, this independent-error model is illustrated geometrically by square boxes around the three predicted patterns.
 
Alternatively, a theory might assume that latent preference states randomly fluctuate across measurement occasions (e.g., across time, persons, or situations), whereas the response process is error-free \citep{regenwetter2017constructbehavior}. 
This means that at each measurement occasion, one of the predicted patterns describes the ``true'' data-generating mechanism perfectly.
However, since we do not know which latent states generated the responses in which trials, this error specification leads to a finite mixture model over the predicted patterns \citep{regenwetter2014qtest}.
Figure~\ref{f.pred}C shows the parameter space of this mixture model for our example.
Essentially, the model permits only those probability vectors $\bm\theta$ that are inside the triangle obtained by connecting the three predicted preference patterns by straight lines (i.e., $\theta_{11} \geq 1 - \theta_{21}$).
Geometrically, this area is the convex hull of the finite number of predicted patterns $\bm v^{(s)}$ and defines a convex polygon in two dimensions (cf. Eq.~\eqref{e.convex} below).
More generally, for $D=3$ choice probabilities, the convex hull results in a convex polyhedron, and for arbitrary number of probabilities $D$, this geometric object is known as a convex polytope \citep{suck1992geometric, koppen1995random}.

The present paper is concerned with mixture models as that illustrated in Figure~\ref{f.pred}C.
Theoretically, these models assume random variation in the latent, data-generating process, which can be represented statistically as a mixture distribution over the finite set of predicted patterns $\bm v^{(s)}$ \citep[][]{regenwetter2017constructbehavior}.
The parameter space of these models can equivalently be described by specifying explicit linear inequality constraints on choice probabilities (e.g., $\theta_{i} \leq \theta_{j}$), or by the convex hull of all response patterns $\bm v^{(s)}$ that are predicted by a theory. 
These mixture models are quite general and, depending upon the experimental design, can provide a strong test of the theory/axiom of interest.  
For example, applied to a single individual with choice responses aggregated over multiple time points, a violation of a mixture model over a set of predictions provides evidence that this individual must have violated the theory of interest; as the model allowed for an arbitrary distribution over all possible theory-consistent preferences.

\section{Multinomial Models with Linear Inequality Constraints}
\label{s.multinomial}

In this section, we outline the notation, likelihood function, and prior distribution of multinomial models and introduce the two equivalent formal representations of linear inequality constraints.

\subsection{Notation and Likelihood}
\label{s.notation}

In the following we use the term ``item type'' to refer to a category system $i$ that is modeled by a multinomial distribution with a fixed total number of observations $n_i$.
For instance, an item type might refer to a certain context, an experimental condition, a pre-existing group, or a specific combination of choice alternatives presented to the participants.
Each item type $i$ has $J_i$ response options indexed by $j=1,\dots,J_i$, and thus the total number of categories is $J = \sum_j J_j$.
This notation is very general, since it includes binary items as a special case (resulting in binomial instead of multinomial distributions), allows us to jointly model items with varying numbers of response options $J_i$ (e.g., binary and ternary items), and even covers paradigms such as ranking tasks.
For instance, if item type $i$ asks participants to rank three elements $\{a,b,c\}$, the six possible rankings $\{abc, acb, bac, bca, cab, cba\}$ define an item type with $J_i=6$ observable response categories.

Since probabilities have to sum to one within each category system, the vector of free parameters $\bm\theta$ is defined by omitting the last probability for each item type,
\begin{equation}
\bm\theta = (\theta_{11},\dots,\theta_{1 (J_1-1)},\theta_{21},\dots, \theta_{I (J_I-1)}).
\end{equation}
This parameter vector contains $D = \sum_i J_i- I$ free parameters for which we will define a likelihood function, specify and test inequality constraints, and derive a Gibbs sampler.
Assuming independent and identically distributed responses within and across item types, the likelihood of the unconstrained model is given as the product of $I$ probability mass functions of the multinomial distribution,
\begin{equation}
\label{e.multinomial}
p(\bm k \mid \bm\theta) = \prod_{i=1}^I  
{n_i \choose k_{i1} ,\dots ,k_{i,J_i} }
\prod_{j=1}^{J_i}  \theta_{ij}^{k_{ij}}
\end{equation}
where the fixed parameters are defined as $\theta_{i,J_i} = 1- \sum_{r=1}^{J_i-1} \theta_{ir}$ for notational convenience.
Note that, in a Bayesian context, it is sufficient to assume only exchangability instead of independent and identically distributed responses \citep{karabatsos2005exchangeable}.
Since the probabilities must sum to one within each multinomial distribution, the unconstrained parameter space $\Omega$ for probability vectors $\bm \theta$ is
\begin{equation}
\Omega = \bigtimes_{i=1}^I \left\{ \bm \theta \in [0,1]^{J_i - 1}  \,\middle|\,  \sum_{j=1}^{J_i - 1} \theta_j  \leq 1 \right\}.
\end{equation}

\subsection{Inequality Constraints}
\label{s.representation}
\textcolor{black}{We are now in a position to connect the likelihood, responsible for generating the data, with the theory being considered. For multinomial models, theories are often operationalized via constraints on the $\bm\theta$ parameters.}
If a psychological theory implies inequality constraints on the probability vector $\bm\theta$ (e.g., $\theta_{ij} \leq \theta_{kl}$), the parameter space of admissible parameters is constrained to a smaller subset $\Omega_c \subset \Omega$.
A model with such a truncated parameter space is more parsimonious than the unconstrained model because it permits a smaller set of probability vectors to account for the data \citep{myung1997applying}.
In the present paper, we are only concerned with inequality constraints that result in a nested parameter space $\Omega_c$ with the same dimensionality as $\Omega$ (cf. Section~\ref{s.future}).
Psychological theories often imply a set of linear inequality constraints, for instance, that a set of binary choice probabilities $\theta_{ij}$ is ordered and increases monotonically across conditions, $\theta_{11} \leq \theta_{21} \leq ...\leq \theta_{I1}$ \citep{heck2017information}.
To summarize a set of $R$ linear inequalities that need to hold jointly, it is convenient to describe the constrained parameter space $\Omega_c$ by a matrix $\bm A \in \mathbb R^{R \times D}$  and a vector $\bm b \in \mathbb R^{R}$ as follows: 
\begin{equation}
\label{e.Ab}
\Omega_c = \left\{\bm\theta \in \Omega \, \middle | \,\bm A \, \bm\theta \leq \bm b \right\},
\end{equation}
where the vector inequality $\bm A \,\bm\theta\leq \bm b$ is defined by element-wise inequalities of the components, that is, $\bm A_{r\bullet} \bm\theta \leq b_r$ must hold for all rows $\bm A_{r\bullet}$. 

\begin{figure}[th]
\centering
\includegraphics[width=10cm]{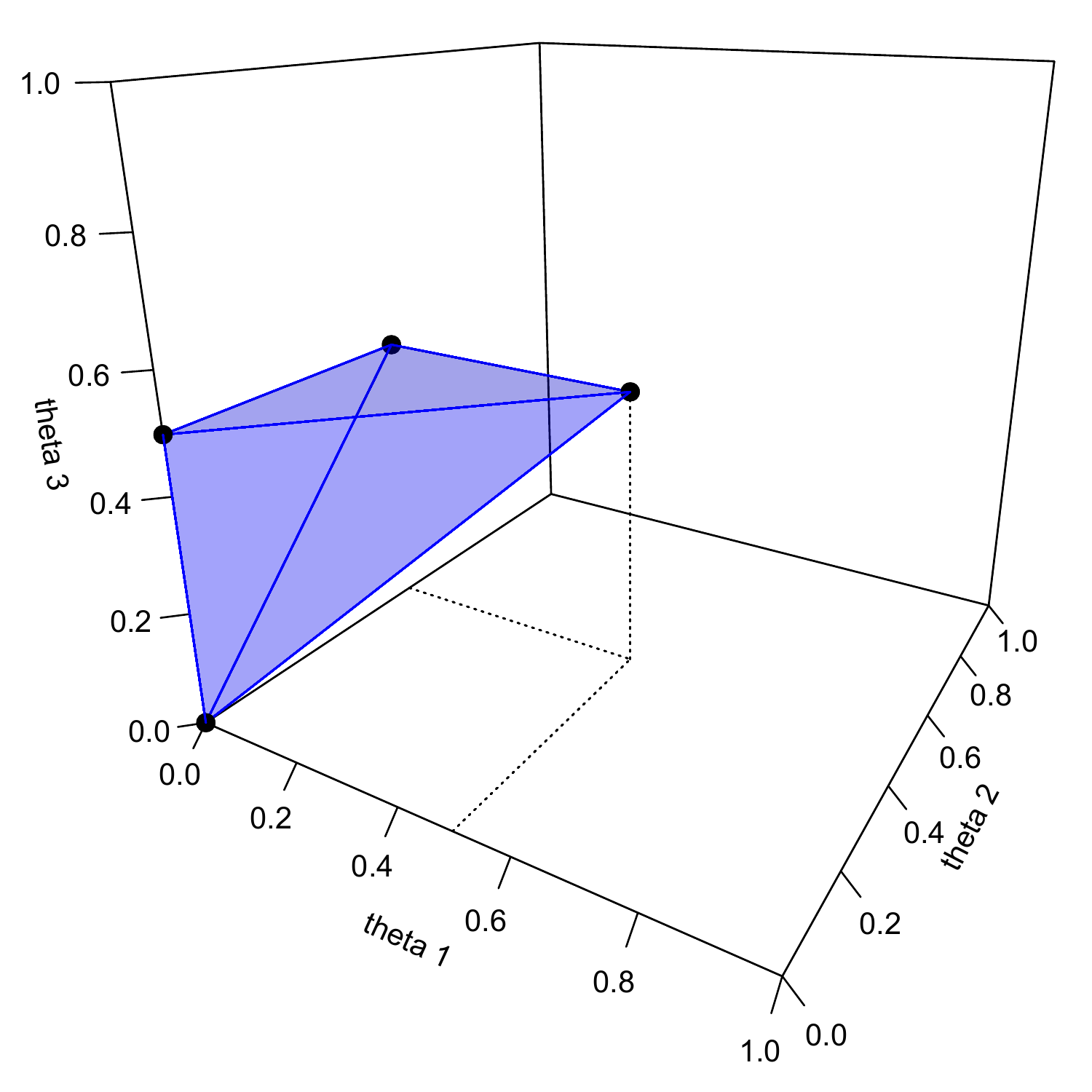}
\caption{A convex polytope can be described either by the facet-defining inequalities as in the $Ab$-representation (i.e., the black sides of the polytope; Eq.~\eqref{e.Ab} and Eq.~\eqref{e.Ab_example}) or by the vertices as in the $V$-representation (i.e., the edges shown by black points; Eq.~\eqref{e.V_example} and Eq.~\eqref{e.V_example2}).}
\label{f.3d}
\end{figure}

As an example, consider the set of monotonic order constraints for binary choice probabilities $\theta_{11}\leq \theta_{21}\leq\theta_{31}\leq .50$ \citep{hilbig2014generalized}, which is described by the $Ab$-representation via
\begin{equation}
\label{e.Ab_example}
\bm A = \begin{pmatrix}
 1 & -1 & 0\\ 
 0 & 1 & -1\\ 
 0 & 0 & 1\\ 
\end{pmatrix} \text{ and }
\bm b = \begin{pmatrix}
   0\\0\\.50
\end{pmatrix}.
\end{equation}
When checking the first row of $\bm A$, we see that $1\cdot \theta_{11} + (-1) \cdot \theta_{21} + 0 \cdot \theta_{31} \leq 0$ is equivalent to the first order constraint $\theta_{11} \leq \theta_{21}$. \textcolor{black}{The second row of $\bm A$ encodes $\theta_{21} \leq \theta_{31}$ and the last row encodes the final order.}
Figure~\ref{f.3d} shows the set of choice probabilities that satisfy these \textcolor{black}{three} inequalities.
In this 3-dimensional parameter space for the parameter vector $\bm\theta=(\theta_{11}, \theta_{21},\theta_{31})$, the inequality constraints are represented by 2-dimensional planes (i.e., by the sides or facets of the polytope). 
If the number of inequalities cannot be reduced any further by omitting redundant constraints, the set of linear inequalities is said to be \emph{facet-defining} \citep{davis-stober2009analysis}.

As an alternative to specifying inequalities on the choice probabilities as in Eq.~\eqref{e.Ab}, the restricted parameter space can equivalently be represented by enumerating the predicted patterns of a theory (cf.~Section~\ref{s.theory}). Geometrically, this representation lists all vertices $\bm v^{(s)}$ of the polytope instead of describing its facets.
For instance, the model defined by the $R=3$ inequalities in Eq.~\eqref{e.Ab_example} can alternatively be specified by $S=4$ vertices.
Graphically, these vertices are the coordinates of the four corners of the 3-dimensional polytope shown by black points in Figure~\ref{f.3d}.
Technically, this means that the truncated parameter space $\Omega_c$ can be defined by the convex hull of the vertices $\bm v^{(s)}$, which contains all possible convex linear combinations of the four vertices \citep{fukuda2004there}:
\begin{equation}
\label{e.V_example}
\bm \theta = 
\alpha_1 \begin{pmatrix} 0\\ 0\\0 \end{pmatrix} +
\alpha_2 \begin{pmatrix} 0\\ 0\\.50 \end{pmatrix} +
\alpha_3 \begin{pmatrix} 0\\ .50\\.50 \end{pmatrix} +
\alpha_4 \begin{pmatrix} .50\\ .50\\.50 \end{pmatrix},
\end{equation}
where $\alpha_1,\dots,\alpha_4$ are (arbitrary) nonnegative weights that sum to one.
Substantively, the mixture weights $\alpha_s$ are simply the probabilities that a specific pattern $\bm v^{(s)}$ generates the observed responses.%
    \footnote{Since the three binomial conditions provide only three degrees of freedom, the four parameters $\bm\alpha$ are not identifiable. Nevertheless, the model is testable \citep{bamber2000how}.}
The restricted parameter space $\Omega_c$ is thus defined as the convex hull of the predicted patterns $\bm v^{(s)}$ \citep{regenwetter2014qtest}:
\begin{equation}
\label{e.convex}
\Omega_c = \left\{\bm \theta = \sum_{s=1}^S \alpha_s \bm v^{(s)} \,\middle |\, \alpha_s \geq 0 \text{ for all } s=1,\dots, S \text{ and }\sum_{s=1}^S \alpha_s=1 \right\}.
\end{equation}
For convenience, we list the vertices in an $S \times D$ matrix $\bm V$, in which each row refers to one vertex $\bm v^{(s)}$ for the free parameters $\bm \theta$:
\begin{equation}
\label{e.V_example2}
\bm V = \begin{pmatrix}
0 & 0 & 0 \\
0 & 0 & .50 \\
0 & .50 & .50 \\
.50 & .50 & .50 \\
\end{pmatrix}.
\end{equation}

Any convex, inequality-constrained model can be described either by the $Ab$-representation using the facet-defining inequalities or by the $V$-representation using a set of vertices to describe the convex hull \citep{davis-stober2009analysis,doignon2016primary,fukuda2004there}. 
Depending on the application, however, one of the two representations is often more convenient for theoretical or technical reasons.
For instance, the $S$ vertices are often easier to derive because they are identical to the predicted patterns $\bm v^{(s)}$ that are implied by a psychological theory (cf. Section~\ref{s.theory}). 
For relatively simple cases (e.g., up to several thousand inequalities), the software \texttt{PORTA} \citep{christof1997porta} or \texttt{polymake} \citep{assarf2017computing} provide algorithms for the conversion of the two representations. 
However, this problem is NP-hard for general convex polytopes, and thus the representation conversion is often infeasible in practice if the dimension $D$ of the polytope or the number of inequalities or vertices is very large. 
For an overview of algorithms for the vertex and facet enumeration problems, see \citet{avis1997how}.

\subsection{Prior and Posterior Distributions}
\label{s.prior}

In the Bayesian framework, a prior distribution is required for the probability parameters $\bm\theta$.
We assume independent Dirichlet distributions for the choice probabilities of the multinomial model, with a truncated support that is defined by the convex, inequality-constrained parameter space $\Omega_c$ \citep{karabatsos2005exchangeable}. 
Given the shape parameters $\bm\beta=(\beta_{11}, \dots, \beta_{IJ})$, the probability density function of the truncated Dirichlet distribution is defined as
\begin{equation}
\label{e.prior}
p(\bm\theta) = \frac {1}{c} \, \mathbb I_{\Omega_c}(\bm\theta) \, 
 \prod_{i=1}^I \prod_{j=1}^{J_i} \theta_{ij}^{\beta_{ij}-1}
\end{equation}
where $\mathbb I_{\Omega_c}(\bm\theta)$ is the indicator function, which equals one if $\bm\theta \in \Omega_c$ and zero otherwise, and $c$ is the normalization constant, which ensures that the density integrates to one.
The special case $\beta_{ij}=1$ for all shape parameters results in a uniform probability distribution, meaning that all of the admissible parameters in $\Omega_c$ are equally likely a priori.
In Section~\ref{s.future}, we discuss the substantive motivation for this prior in more detail.

The normalizing constant $c$ of the prior distribution is obtained by integrating the kernel\footnote{\textcolor{black}{To clarify, we use the term ``kernel'' to refer to the form of the probability density function where any factors that are not functions of variables are removed.}} of the prior probability density in Eq.~\eqref{e.prior} over the the restricted parameter space $\Omega_c$:
\begin{equation}
\label{e.constant_c}
c =  \int_{\Omega_c} \prod_{i=1}^I \prod_{j=1}^{J_i} \theta_{ij}^{\beta_{ij}-1} \diff \bm\theta.
\end{equation}
Since $c$ necessarily decreases as the relative volume of $\Omega_c$ becomes smaller, this constant quantifies the restrictiveness or parsimony of the inequality-constrained model.
For some types of order constraints and prior parameters, the constant $c$ can be derived analytically.
This is often the case when assuming a uniform prior distribution on the choice probabilities (i.e., $\beta_{ij}=1$) because then, the integral in Eq.~\eqref{e.constant_c} simply equals the volume of the restricted parameter space $\Omega_c$.
For example, consider the inequality constraints $\theta_{11}\leq \theta_{21}\leq\theta_{31}\leq .50$ illustrated in Figure~\ref{f.3d}.
In this case, the linear order on the three parameters divides the volume of the 3-dimensional cube with side length $0.50$ (and thus, a volume of $0.50^3$) into $3!$ equally-sized parts, which results in 
\begin{equation}
c = 0.50^3 \cdot \frac{1}{3!}.
\end{equation}
However, even when assuming a uniform prior, an analytical solution for $c$ is often difficult to find because a large number of facets or vertices often results in very complex polytopes for the parameter space $\Omega_c$.
As a remedy, the constant $c$ often needs to be estimated by Monte Carlo integration (cf. Section~\ref{s.Ab_bf}).

Since the Dirichlet distribution is a conjugate prior for the multinomial distribution, the posterior distribution is also a Dirichlet \citep{lindley1964bayesian},
\begin{equation}
\label{e.posterior}
p(\bm\theta \mid \bm k) = \frac {1}{f} \, \mathbb I_{\Omega_c}(\bm\theta) \, 
\prod_{i=1}^I \prod_{j=1}^{J_i} \theta_{ij}^{k_{ij} + \beta_{ij}-1}.
\end{equation}
The normalizing constant $f$ of this truncated distribution is computed via integration similarly as for the prior in Eq.~\eqref{e.constant_c} while replacing the exponents in the integrand by $k_{ij} + \beta_{ij}-1$.
Note that the prior shape parameters $\beta_{ij}$ additively combine with the observed frequencies $k_{ij}$ and can thus be interpreted as the prior sample size assigned to each response category.

\section{Bayesian Inference Using the Inequality Representation}
\label{s.Ab}

In this section, we summarize and improve computational methods for the Bayesian analysis of multinomial models given a set of linear inequality constraints.

\subsection{Gibbs Sampling for the $Ab$-Representation}
\label{s.Ab_gibbs}

In the Bayesian framework, parameter estimation focuses on the posterior distribution of the parameters given the data.
Usually, computational methods are required to obtain point and uncertainty estimates for the parameters.
For this purpose, Markov chain Monte Carlo (MCMC) methods draw samples from the posterior distribution, which can then be summarized by descriptive statistics such as the mean, standard deviation, or highest-density intervals.

The Gibbs sampler is a specific MCMC algorithm that cycles through the components of a parameter vector by drawing samples of the conditional posterior distributions of one parameter given the remaining parameters.
Gibbs samplers are especially useful for models with inequality-constraints, because the constraints merely truncate the range of admissible parameter values, wheres the shape or kernel of the posterior density function remains proportional to that of the unconstrained model \citep{gelfand1992bayesian}.
In previous work, Gibbs samplers have been developed for binomial models with specific, theoretically-derived sets of inequality constraints.
For instance, \citet{myung2005bayesian} constructed a sampler tailored to decision axioms such as weak or strong stochastic transitivity, \citet{karabatsos2004orderconstrained} derived constraints based on item response theory, and \citet{prince2012design} developed a Gibbs sampler for state-trace analysis.
We generalize these approaches by developing a Gibbs sampler for multinomial data given any set of convex, linear inequality constraints defined by the inequality representation $\bm A \, \bm\theta \leq \bm b$.
By relying on an analytical solution of the conditional posterior distribution, the proposed Gibbs sampler is more efficient compared to MCMC samplers that require accept-reject steps in each iteration \citep[e.g., Metropolis within Gibbs;][]{karabatsos2001rasch}.
Our approach also generalizes the hit-and-run-sampler \citep{smith1984efficient}, an efficient MCMC method that draws random samples from a uniform distribution on a convex polytope \citep{lovasz2006hitandrun}.

\begin{figure}[th]
\centering
\includegraphics[width=10cm]{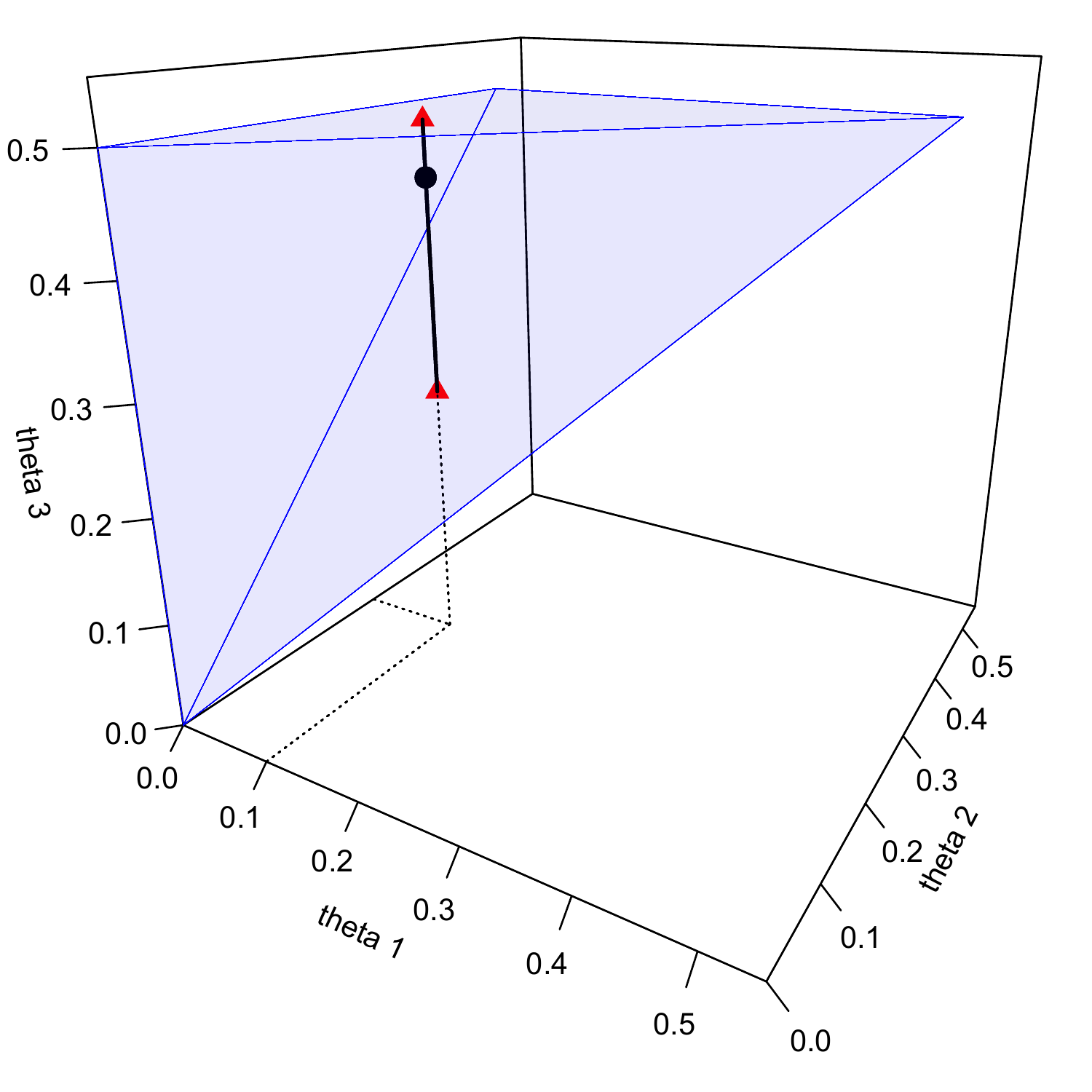}
\caption{
Given a specific parameter vector $\bm\theta^{(t)}$ within the polytope (black point), the Gibbs sampler ``walks'' in one of the coordinate directions (here, the vertical direction $i=3$ shown by a solid black line).
To compute the support for the next random sample $\theta_{i1}^{(t+1)}$, the intersection of this direction vector with the polytope is computed (red triangles).
}
\label{f.conditional}
\end{figure}

In each iteration $t$, the Gibbs sampler cycles through the elements $\theta^{(t)}_{ij}$ of the parameter vector $\bm\theta^{(t)}$ (either sequentially or at random) and updates the current probability parameter $\theta^{(t)}_{ij}$ given the remaining parameters.
This updating step requires the conditional posterior distribution of the probability parameter $\theta_{ij}$ for item type $i$ and response option $j$ \citep{gelfand1992bayesian}. 
To derive this distribution, we first consider the support of the parameter $\theta_{ij}$ conditional on the remaining parameters $\bm\theta_{-ij}=(\theta_{11},\dots,\theta_{i(j-1)},\theta_{i(j+1)}, ,\dots,\theta_{I,J_I})$.
Geometrically, this problem is identical to computing the intersection of a line with a convex polytope as illustrated in Figure~\ref{f.conditional}.
This problem is also known as ``line-clipping'' for 3-dimensional polyhedra, and thus the following solution builds on ideas of the Cyrus-Beck algorithm often used in computer graphics \citep{cyrus1978generalized}.
Given a specific parameter $\bm\theta^{(t)}$ inside the polytope (black point), the method searches for the lower an upper bounds in the direction of the parameter dimension of $\theta_{ij}$ while fixing the remaining parameters $\bm\theta_{-ij}$ (shown by the vertical black line). 
By solving the set of inequalities as derived below, we obtain the lower and upper truncation boundaries (red triangles).

To derive the lower and upper truncation boundaries for the parameter $\theta_{ij}$, we denote the $ij$-th column vector of $\bm A$ by $\bm A_{\bullet (ij)}$ and the remaining matrix without this column by $\bm A_{\bullet (-ij)}$.
Using vector notation, we split the sum on the left side of the system of inequalities $\bm A \,\bm \theta \leq \bm b$ into two parts,
\begin{equation}
\theta_{ij} \bm A_{\bullet (ij)} +  \bm A_{\bullet (-ij)} \bm \theta_{-ij}\leq \bm b.
\end{equation}
Next, we subtract all terms involving the remaining parameters $\bm\theta_{-ij}$,
\begin{equation}
\label{e.Ab_split}
\theta_{ij} \bm A_{\bullet (ij)} \leq \bm b - \bm A_{\bullet (-ij)} \bm \theta_{-ij}
\end{equation}
Finally, the lower and upper truncation limits $z_0$ and $z_1$ are derived by dividing both sides by the column entries $\bm A_{\bullet (ij)}$, while reversing those inequalities for which the components $A_{r\,(ij)}$ in the $r$-th row and $ij$-th column of $\bm A$ are negative,
\begin{equation}
\label{e.Ab_bounds}
\begin{cases}
\theta_{ij}\geq z_0 := \max \{(b_r - \bm A_{r \,(-ij)}\bm\theta_{-ij}) / A_{r \, (ij)} \text{ for all } r \text{ with } A_{r \, (ij)}<0\}\\
\theta_{ij}\leq z_1 := \min \{(b_r - \bm A_{r \,(-ij)}\bm\theta_{-ij}) / A_{r \, (ij)} \text{ for all } r \text{ with }A_{r \, (ij)}>0\},\\
\end{cases}
\end{equation}
where $\bm A_{r\,(-ij)}$ is the $r$-th row vector of $\bm A$ without the $ij$-column entry.
In Eq.~\eqref{e.Ab_bounds}, we compute the maximum (minimum) of all lower-bound (upper-bound) inequalities since the parameter $\theta_{ij}$ must satisfy \emph{all} inequalities defined by the vector notation in Eq.~\eqref{e.Ab_split}.
Overall, this derivation shows that the conditional posterior of $\theta_{ij}$ has support on the interval $[z_0, z_1]$. \textcolor{black}{This follows via convexity of the constrained parameter space.}

Next, we derive the kernel density of the conditional posterior on the support $[z_0, z_1]$.
For this purpose, we define the scaling factor
\begin{equation}
s_{ij}  = 1 - \sum_{p=1, p \neq j}^{J_i - 1} \theta_{ip} = \theta_{ij} + \theta_{i,J_i}
\end{equation}
which equals the sum of the probabilities for the choice frequencies $k_{ij}$ and $k_{i,J_i}$.
Using this notation, it follows that the conditional posterior distribution of $\theta_{ij}$ given fixed values of the remaining probabilities $\bm\theta_{-ij}$ is 
\begin{align}
f(\theta_{ij} \mid \bm\theta_{-ij}, \bm k) 
&\propto  \theta_{ij}^{k_{ij}} \left(1 - \sum_{j=1}^{J_i - 1}\theta_{ij}\right)^{k_{J_i}} \mathbb I_{[z_0, z_1]}(\theta_{ij})\\
&\propto  \theta_{ij}^{k_{ij}} \left(s_{ij}  - \theta_{ij}\right)^{k_{J_i}} \mathbb I_{[z_0, z_1]}(\theta_{ij})\\
&\propto \left(\theta_{ij} / s_{ij} \right)^{k_{ij}} 
\left(1- \theta_{ij}/s_{ij}\right)^{k_{i,J_i}} \mathbb I_{[z_0/s_{ij}, z_1/s_{ij}]}(\theta_{ij}/s_{ij}). \label{e.conditional}
\end{align}

Importantly, with respect to the scaled parameter $\theta_{ij} / s_{ij}$, Eq.~\eqref{e.conditional} is proportional to the density function of a truncated beta distribution:
\begin{equation}
\label{e.conditional_stbeta}
(\theta_{ij} / s_{ij} \mid \bm\theta_{-ij}, \bm k) \sim \text{Beta}(k_{ij} + \beta_{ij}, k_{i,J_i}+\beta_{i,J_i} ) \text{ truncated to } [z_0/s_{ij} , z_1/s_{ij}].
\end{equation}
This analytical solution allows us to sample from the conditional posterior distribution of $\theta_{ij}$ efficiently.
First, one draws a (not yet scaled) random sample $\eta_{ij}^{(t)}$ from a truncated beta distribution using the  inverse transformation method \citep[][p. 38]{devroye1986nonuniform},
\begin{equation}
\eta^{(t)}_{ij} = F^{-1}\left[F(z_0) + \left( F(z_1) - F(z_0)\right) u^{(t)}\right],
\end{equation}
where $u^{(t)}$ is sampled uniformly on the interval $[0,1]$ and $F$ is the cumulative density function of the non-truncated beta distribution with shape parameters defined in Eq.~\eqref{e.conditional_stbeta}.
In the next step, the sampled value $\eta^{(t)}_{ij}$ is scaled by setting $\theta_{ij}^{(t)} = s_{ij} \eta_{ij}^{(t)}$, which produces a sample from the conditional target distribution in Eq.~\eqref{e.conditional}.

By using this analytical solution for the conditional posterior distribution, we can then update the current values of the parameter vector $\bm\theta^{(t)}$ either in fixed or random order \citep[systematic vs. random scan;][p.~375]{robert2004monte} within each of the $T$ iterations of the Gibbs sampler.
Since the Gibbs sampler requires a point inside the restricted parameter space $\Omega_c$ as a starting value, we use convex-constrained optimization to find the maximum a posteriori (MAP) estimate of $\bm\theta$ \citep{lange2010numerical}.  
If a uniform prior is assumed (i.e., all $\beta_{ij}=1$), the posterior distribution is guaranteed to be uni-modal.
Thus, using the MAP estimate as a starting value improves convergence of the Gibbs sampler and reduces the number of required burn-in samples.

\subsection{Posterior-Predictive $p$-Values}
\label{s.ppp}

In statistical modeling, it is often important to test the absolute fit of a model to data.
In the Bayesian framework, posterior-predictive $p$-values provide a measure of fit that is both intuitive and easy to compute \citep{meng1994posterior}.
To quantify the discrepancy between observed and expected frequencies, it is convenient to rely on Pearson's $X^2$-statistic for multinomial models,
\begin{equation}
\label{e.X2}
X^2 = \sum_{i=1}^{I} \sum_{j=1}^{J_i} \frac{\left(k_{ij} - \theta_{ij}n_i \right)^2}{\theta_{ij}n_i}.
\end{equation}
Essentially, posterior-predictive $p$-values compare the distribution of the $X^2$-statistic for the observed frequencies $k_{ij}$ against that for the posterior-predictive frequencies $k_{ij}^\text{(t)}$.

From a practical perspective, posterior-predictive $p$-values are computed by iterating through all $T$ posterior samples.
In each iteration, we apply Eq.~\eqref{e.X2} to compute 
(a) the statistic $X_\text{obs}^{2\;(t)}$ using the observed frequencies $k_{ij}$ and 
(b) the statistic $X_\text{pred}^{2\;(t)}$ using the posterior-predictive frequencies $k^{(t)}_{ij}$, which are randomly sampled from the product-multinomial distribution in Eq.~\eqref{e.multinomial} based on the posterior samples $\bm\theta^{(t)}$.
In both cases, the expected frequencies are $\theta_{ij}^{(t)}n_{i}$.
Finally, the posterior-predictive $p$-value is computed as the proportion of $X^2$-samples for which the observed test statistic is smaller than the posterior-predictive one,
\begin{equation}
p_{B} = \frac 1 T \sum_{t=1}^T 
\mathbb I_{\{X_\text{obs}^{2\;(t)} < X_\text{pred}^{2\;(t)}\}}. 
\end{equation}
This Bayesian $p$-value will be very small if the observed frequencies do not match the distribution of frequencies implied by the posterior distribution of the model.
However, even though small $p_B$ indicate misfit, their interpretation differs from that of $p$-values in classical statistics because posterior-predictive $p$-values are not uniformly distributed under the null hypothesis \citep[][]{meng1994posterior}.

Note that different test statistics can be used to obtain posterior-predictive $p$-values tailored to specific research questions.
For instance, one can compute separate $X^2$-statistics for different experimental conditions or stimuli to test whether model fit is moderated by these factors.
Another advantage of posterior-predictive $p$-values is that their computation requires only posterior samples, which can efficiently be obtained using the Gibbs sampler.
In contrast, computation of the Bayes factor may often be computationally more costly especially for complex inequality constraints.

\subsection{The Encompassing Bayes Factor for the $Ab$-Representation}
\label{s.Ab_bf}
 
To test whether the predictions of a theory are valid, we compare the model $\mathcal M_u$ with the unconstrained parameter space $\Omega$ against the model $\mathcal M_c$ with the inequality-constrained parameter space $\Omega_c$.
Since the constrained model is a nested model, it can never fit the data better than the unconstrained model.
However, due to the restricted parameter space, the constrained model $\mathcal M_c$  is more parsimonious, a property that is desirable from a theoretical perspective because it results in improved prediction accuracy for new data \citep{myung1997applying}.
To compare different models, Bayesian model selection provides a trade-off between model fit and complexity. 
More specifically, the Bayes factor $B_{cu}$ quantifies the evidence for the inequality-constrained model $\mathcal M_c$ versus the unconstrained model $\mathcal M_u$ and is defined as the ratio of the marginal likelihoods of the two models:
\begin{equation}
\label{e.bf}
B_{cu} := \frac {p(\bm k \mid \mathcal M_c) }{p(\bm k \mid  \mathcal M_u)}
= \frac
{\int_{\Omega_c} p(\bm k \mid \bm\theta, \mathcal M_c) p(\bm\theta \mid \mathcal M_c) \diff \bm\theta }
{\int_{\Omega}p(\bm k \mid \bm\theta, \mathcal M_u) p(\bm\theta \mid \mathcal M_u) \diff \bm\theta }.
\end{equation}

Usually, it is difficult to compute the integrals in Eq.~\eqref{e.bf} directly.
As a remedy, we rely on the method of encompassing Bayes factors to compute the Bayes factor \citep{klugkist2005bayesian, klugkist2007bayes}.
Within this framework, the prior of the nested model $\mathcal M_c$ must be proportional to the prior of the encompassing model $\mathcal M_u$ within the constrained parameter space $\Omega_c$.
Technically, the priors thus have the same kernel density and only differ by the support and the normalizing constant.
The truncated Dirichlet prior in Eq.~\eqref{e.prior} satisfies this requirement, because the nested and the unconstrained model differ only in the parameter spaces $\Omega_c$ versus $\Omega$, respectively.
To facilitate the computation of the Bayes factor, \citet{klugkist2005bayesian} used the well-known representation of the marginal probability $p(\bm k) = p(\bm k \mid \bm\theta^*) p(\bm \theta^*)/ p(\bm \theta^* \mid\bm k)$  for an arbitrary parameter value $\bm\theta^* \in \Omega_c$. 
When applied to Eq.~\eqref{e.bf}, we obtain the following identity: 
\begin{align}
B_{cu} &:= \frac
{p(\bm k \mid \bm\theta^*, \mathcal M_c) p(\bm\theta^* \mid \mathcal M_c) / p(\bm\theta^* \mid \bm k, \mathcal M_c)  }
{p(\bm k \mid \bm\theta^*, \mathcal M_u) p(\bm\theta^* \mid \mathcal M_u) / p(\bm\theta^* \mid \bm k, \mathcal M_u)} \nonumber \\
&= \frac
{\int_{\Omega_c} p(\bm\theta \mid \bm k, \mathcal M_u) \diff \bm\theta }
{\int_{\Omega_c} p(\bm\theta \mid \mathcal M_u) \diff \bm\theta }
= \frac{f}{c}.  \label{e.bf_encomp}
\end{align}
This derivation uses the fact that for $\bm\theta^*\in \Omega_c$, the likelihood function $p(\bm k \mid \bm\theta^*, \mathcal M)$ is identical for the two models, and the prior $p(\bm\theta^* \mid \mathcal M)$ and the posterior $p(\bm\theta^* \mid \bm k, \mathcal M)$ are proportional up to the constants $c$ and $f$, respectively.
Hence, it follows that the Bayes factor equals the ratio of the normalizing constants of the prior and posterior distribution for the \emph{constrained} model $\mathcal M_c$ (see Eq.~\eqref{e.prior} and Eq.~\eqref{e.posterior}).


To approximate the constants $c$ and $f$ in Eq.~\eqref{e.bf_encomp}, we can use Monte Carlo sampling to estimate the proportion of prior and posterior samples from the \emph{encompassing} model that satisfies the constraints.
More precisely, computing the encompassing Bayes factor requires the following steps \citep[for a detailed introduction and proofs, see][]{hoijtink2011informative,hoijtink2008bayesian}.
First, we draw $T$ random samples from the prior distribution of the unconstrained model $\mathcal M_u$, which can easily be done for the Dirichlet distribution.
Second, we count the number of samples $\bm\theta^{(t)}$ that are within the parameter space of the constrained model $\mathcal M_c$.
Given the $Ab$-representation of an inequality-constrained multinomial model, we only have to check\footnote{
	To increase computational efficiency, we implemented this check as a while-loop, which stops once a single constraint is violated.
	Efficiency can further be increased by ordering the $Ab$-inequalities according to the number of violations that occur during sampling (since less checks are required if frequently-violated inequalities are tested first).} whether $\bm A \, \bm\theta^{(t)} \leq \bm b$.
The observed proportion of prior samples in $\Omega_c$ is denoted by $\hat c$ and provides an estimate for the constant $c$.
Next, we draw random samples from the posterior distribution of the unconstrained model $\mathcal M_u$, which is also straightforward because the posterior is a conjugate Dirichlet distribution.
Similar as for the prior samples, we compute the proportion of posterior samples $\hat f$ that are within the constrained parameter space $\Omega_c$ as an approximation for the normalizing constant $f$. 

Importantly, the encompassing Bayes factor requires prior and posterior samples from the \emph{encompassing} model only.
Thereby, the approach is closely related to the popular Savage-Dickey density ratio for computing the Bayes factor in favor of an \emph{equality} constraint \citep{wetzels2010encompassing, heck2018caveat}. 
Recently, the method of encompassing priors has been implemented for binomial models in the software \textsc{QTest} \citep{regenwetter2014qtest, regenwetter2019tutorial}.

\subsection{Precision of the Encompassing Bayes Factor Approximation}
\label{s.bf_SE}

Despite the advantages of being computationally straightforward, the encompassing Bayes-factor approach only provides an approximation of the Bayes factor $\hat B_{cu} = \hat f / \hat c$.
In substantive applications, it is important to ensure that the Bayes factor approximation is sufficiently precise to draw any substantive conclusions.
In the following, we show how to quantify the uncertainty based on recommendations by \citet{hoijtink2011informative}.
Importantly, this approach quantifies the uncertainty of the approximation due to the specific \emph{computational} implementation \citep[for a similar approach, see][]{heck2018quantifying} and does not provide any information about the \emph{empirical} sampling variation of the Bayes factor for new data.

The precision of the Bayes factor approximation in Eq.~\eqref{e.bf_encomp} can be quantified by formalizing the sampling process used to approximate the constants $c$ and $f$ \citep{hoijtink2011informative}.
Given that $T$ posterior samples were drawn to approximate $f$, the number of samples $T_f$ that satisfy the order constraints can be understood as a binomial random variable with the unknown rate parameter $f$.
To estimate $f$, it is sufficient to compute the maximum-likelihood estimate $\hat f = T_f/T$ as discussed above. 
However, by treating the number of samples $T_f$ as a random variable, we can derive a posterior distribution for the unknown constant $f$.
By assuming a uniform prior, the posterior distribution of $f$ is the beta distribution
\begin{equation}
\label{e.uncertainty_f}
(f \mid T_f) \sim \text{Beta}(T_f + 1, T - T_f + 1).
\end{equation}
This distribution describes the uncertainty associated with the approximation of the constant $f$ given that $T_f$ out of $T$ posterior samples from the encompassing model satisfied the constraints (a similar approach applies to the normalizing constant $c$, which is approximated by the proportion $\hat c$ of \emph{prior} samples satisfying the constraints).

To quantify the uncertainty in the Bayes-factor approximation $\hat B_{cu} = \hat f/\hat c$, we use a sampling approach \citep{hoijtink2011informative}.
That is, we draw $r=1,\dots,R$ samples for the unknown parameters $f$ and $c$ from the beta posterior distributions in Eq.~\eqref{e.uncertainty_f} and compute the Bayes factor as the ratio of these posterior samples:
\begin{equation}
\label{e.uncertainty_bf}
B^{(r)}_{cu} = f^{(r)} /c^{(r)}.
\end{equation}
Thereby, we obtain a distribution of Bayes factors, which quantifies the uncertainty of the numerical approximation \citep{heck2018quantifying}.
Specifically, we can compute the standard deviation of the samples $B^{(r)}_{cu}$ to quantify the precision of $\hat B_{cu} = \hat f/\hat c$.
When testing the same model for multiple participants or data sets, computational time can be saved by approximating the prior constant $c$ once with very high precision (large $T$), whereas the constant $f$ needs to be approximated separately for different vectors of observed frequencies $\bm k$.
Moreover, if the model assumes a uniform prior and if the volume of the parameter space $\Omega_c$ is available in closed form  \citep[e.g.,][]{prince2012design, heck2017information}, the exact numerical value for $c$ can be used in the sampling approximation in Eq.~\eqref{e.uncertainty_bf}.

\subsection{A Stepwise Algorithm}

The encompassing Bayes factor approach has the drawback that very large numbers of samples are required if the constants $c$ and $f$ are very small \citep{hoijtink2011informative}.
This will be the case if the polytope defined by the parameter space $\Omega_c$ has a very small volume relative to the encompassing parameter space $\Omega$.
Especially for $f$, this issue also arises if the posterior distribution of the encompassing model assigns only very small probability mass to the constrained parameter space $\Omega_c$.
In both cases, only a very small proportion of the $T$ samples will be inside $\Omega_c$, and thus, the estimates $\hat c=T_c/T$ and $\hat f=T_f/T$ will have large sampling error.
Unfortunately, this issue becomes even more severe for the Bayes factor approximation, since the total number of samples $T$ cancels out, $\hat B_{cu} = T_c/T_f$.
In the worst case, both $T_c$ and $T_f$ equal zero, in which case the Bayes factor cannot be estimated at all.

As a remedy, \citet{hoijtink2011informative} proposed to split large sets of inequality constraints into monotonically increasing subsets \citep[see also][]{mulder2012biems}, an approach that has not yet been applied or implemented for multinomial models yet.
For the specific scenario of the $Ab$-representation, this can easily be achieved by partitioning the rows of the matrix $\bm A$ and the corresponding entries of the vector $\bm b$. 
For the example in Eq.~\eqref{e.Ab_example}, we can define two nested, inequality-constrained models $\mathcal M_1$ and $\mathcal M_2$ as follows:
\begin{align}
\label{e.Ab_split1}
\mathcal M_1&: \bm A^{(1)} = 
\begin{pmatrix}
 1 & -1 & 0\\ 
\end{pmatrix} \text{ and }
\bm b^{(1)} = \begin{pmatrix} 0\\ \end{pmatrix},\\ \label{e.Ab_split2}
\mathcal M_2&: \bm A^{(2)} = 
\begin{pmatrix}
 1 & -1 & 0\\ 
 0 &  1 & -1\\ 
\end{pmatrix} \text{ and }
\bm b^{(2)} = \begin{pmatrix}  0\\ 0\\ \end{pmatrix}.
\end{align}
Here, the model $\mathcal M_1$ ($\mathcal M_2$) is obtained by selecting the first one (two) rows of $\bm A$ and the first one (two) entries of $\bm b$.
By dropping distinct subsets of the inequality constraints, the parameter space increases monotonically and we obtain an order of nested models, $\mathcal M_c \subset \mathcal M_2 \subset \mathcal M_1 \subset \mathcal M_u$.

Based on this decomposition into an order of nested models, we can now compute the encompassing Bayes factor using a stepwise algorithm.
This approach relies on approximating multiple constants $c_m$ for each pairwise comparison of two nested models.
First, we use prior samples from the Dirichlet distribution of the encompassing model to compute the proportion of samples $\hat c_1$ that are inside the parameter space of $\mathcal M_1$, similar as before.
Second, we use the Gibbs sampler from Section~\ref{s.Ab_gibbs} to draw prior samples from the inequality-constrained model $\mathcal M_1$ and count the proportion of samples $\hat c_2$ that are in $\mathcal M_2$.
Third, we again use Gibbs sampling for the model $\mathcal M_2$ and count the proportion of samples $\hat c_c$ in $\mathcal M_c$.
Since each of the intermediate constants $c_m$ represents the relative decrease in volume for two consecutive, nested models,%
    \footnote{The constant $c_m$ equals the relative volume of the polytope only if the prior is uniform ($\beta_{ij}=1$).  For non-uniform priors, $c_m$ equals the integral $\int_{\Omega_m} p(\bm\theta \mid \mathcal M_{m-1}) \diff \bm\theta$.} 
the overall constant $c$ can be obtained by multiplication \citep[for a detailed proof, see][]{hoijtink2011informative}: 
\begin{equation}
c = c_1 \, c_2 \, c_c.
\end{equation}
The same strategy applies to the posterior constant $f$, with the difference that, in each step of the algorithm, samples are drawn from the truncated posterior instead of the prior distribution.
Finally, due to transitivity of the Bayes factor, the approximation of the encompassing Bayes factor is: 
\begin{equation}
\hat B_{cu} = \hat B_{c1} \, \hat B_{12} \, \hat B_{2u} \approx 
\frac{\hat c_c}{\hat f_c} \, \frac{\hat c_2}{\hat f_2} \, \frac{\hat c_1}{\hat f_1}.
\end{equation}

Why is it that the stepwise approach results in a more precise approximation of the Bayes factor?
Essentially, this is due to the fact that the parameter spaces of two consecutive, nested models $\mathcal M_m$  and $\mathcal M_{m+1}$  differ much less in volume in comparison to the difference in the parameter spaces of the encompassing model $\mathcal M_u$ and the most strongly constrained model $\mathcal M_c$ \citep{hoijtink2011informative}.
Accordingly, each of the constants $c_m$ will be much larger than the overall constant $c$, and can thus be approximated with higher precision.
This in turn increases the precision of the approximation $\hat c$, and in turn, the precision of the Bayes factor estimate $\hat B_{cu} = \hat c/\hat f$. 
To quantify the uncertainty of the stepwise procedure, we can extend the approach from Section~\ref{s.bf_SE}.
Similar as before, we use random samples  $c_m^{(r)}$ from beta distributions to approximate the uncertainty of each intermediate constant in isolation.
Next, we use these samples to repeatedly compute the overall constant $c^{(r)}= c^{(r)}_1 \, c^{(r)}_2 \cdots c^{(r)}_c$.
To summarize the uncertainty of the approximation $\hat c$, we can then compute the standard deviation of the samples $c^{(r)}$.
Moreover, it is possible to use samples  $c^{(r)}$ and  $f^{(r)}$ for both the prior and the posterior constants to judge the uncertainty of the Bayes factor approximation using Eq.~\eqref{e.uncertainty_bf}.

Several details of the stepwise algorithm can be improved to increase the computational efficiency even further.
First, when checking whether samples from an inequality-constrained model $\mathcal M_m$ are inside the parameter space of $\mathcal M_{m+1}$, it is sufficient to check only those constraints that are unique to the model $\mathcal M_{m+1}$ (since we sample from $\mathcal M_m$, we do not need to check whether the corresponding inequalities hold).
Second, to obtain a starting value for the Gibbs sampler of an inequality-constrained model $\mathcal M_m$, we reuse the last sample from the next larger model $\mathcal M_{m-1}$ that satisfied the additional constraints in $\mathcal M_m$. 
Thereby, the Gibbs sampler does not require a burn-in phase for each of the consecutive steps.
Finally, since independent sampling from the encompassing model is more efficient than Gibbs sampling, we recommend to split the inequalities $\bm A \, \bm \theta \leq \bm b$ into larger subsets.
For instance, instead of defining consecutive, nested models by dropping one inequality at a time, the first step from the encompassing model $\mathcal M_u$ to the nested model $\mathcal M_1$ can include multiple inequalities (e.g., dozens) instead of a single inequality.
Thereby, more samples can be drawn from the encompassing model to compute $\hat c_1$ in the first step, whereas fewer samples are sufficient to compute $\hat c_m$ in the remaining Gibbs-sampling steps.

\subsection{An Automatic Stepwise Algorithm}
\label{s.automatic}

Despite the increased precision of the stepwise procedure, the question remains how many samples $T_m$ for each step $m$ are sufficient.
Whereas some steps will require only few samples (e.g., if the prior constant $c_m$ is close to one), others will require more samples to ensure the same level of precision (e.g., if the posterior constant $f_m$ is very small because the constraints in one step are violated by the data).
Especially for the approximation of the posterior constant $f$, it is difficult to judge a-priori which of the steps require more samples.

As a remedy, we propose an automatic stepwise procedure.
For each step $m$, this method continuously samples from the model $\mathcal M_m$  until a minimum number of samples $T_\text{min}>0$ has been observed within the constrained parameter space of the next smaller model $\mathcal M_{m+1}$.
Thereby, more samples are drawn to approximate those constants $c_m$ close to zero which would otherwise have a larger approximation uncertainty for identical $T_m$.
Moreover, this approach also ensures that each of the intermediate approximations $\hat c_m$ are strictly positive, which resolves the issue that the Bayes factor cannot be computed if both $\hat f$ and $\hat c$ are zero.

To ensure that a minimum number of samples is used in each step, we first iterate through all models by drawing $T_{0}$ samples. 
After the first round, we switch between the different nested models and always update the model with the smallest number of samples $T_m$ satisfying the corresponding inequality constraints.
Moreover, as a starting value for the Gibbs sampling from a model $\mathcal M_m$, we again use an adaptive scheme that selects the most recent parameter vector that satisfies the corresponding constraints.
Thereby, we reduce the issue of requiring a burn-in phase in each step.

The uncertainty of the automatic stepwise procedure can again be quantified by drawing samples from beta distributions similar as for the stepwise procedure in the previous section.
This is the case because in each iteration $m$ of the automatic procedure, the sampling process results in a negative-binomial likelihood for the number of samples $T_m$ that is required to reach the minimum number of ``hits'' $T_\text{min}$.
Since the beta distribution is a conjugate prior for the negative binomial, we can again draw samples from a beta distribution (with shape parameters $T_\text{min}+1$ and $T_m-T_\text{min}+1$) to quantify the precision of the proportion $\hat c_m = T_\text{min}/T_m$.

\section{Bayesian Inference Using the Vertex Representation}
\label{s.V}

In the following, we develop computational tools for obtaining posterior samples and computing the Bayes factor for inequality-constrained multinomial models that are defined by the $V$-representation.
Instead of providing a set of inequalities as in the $Ab$-representation, the $V$-representation uses an $S \times D$ matrix that contains one vertex $\bm v^{(s)}$ (e.g., a predicted pattern) per row as illustrated in Eq.~\eqref{e.convex}.
For many psychological theories, it is indeed easier to obtain a list of all admissible predicted patterns \citep{regenwetter2017constructbehavior}.
Since transformation between the two types of representations is in general NP-hard and often infeasible \citep{doignon2016primary, fukuda2004there}, the following developments facilitate the statistical test of psychological theories in practical applications.

\subsection{Gibbs Sampling for the $V$-Representation}
\label{s.V_gibbs}

In Section~\ref{s.Ab_gibbs}, we developed a Gibbs sampler for inequality-constrained multinomial models by deriving the conditional posterior distribution of a parameter $\theta_{ij}$ given the remaining parameters $\bm\theta_{-ij}$.
The same steps are required for the $V$-representation.
However, the posterior distribution of an inequality-constrained model does not depend on the type of representation that is used to define the restricted parameter space $\Omega_c$.
Hence, it follows that both the full posterior distribution and the conditional posterior distributions for the $V$-representation are identical to those derived for the $Ab$-representation in Section~\ref{s.Ab_gibbs}.
Specifically, the conditional posterior of a parameter $\theta_{ij}$ is again the scaled, truncated beta distribution in Eq.~\eqref{e.conditional}.
However, to draw random samples from this distribution, it is necessary to compute the lower and upper truncation boundaries $z_0$ and $z_1$ conditional on the remaining parameters $\bm\theta_{-ij}^{(t)}$.
For the $Ab$-representation, these boundaries were simply derived by solving the set of inequalities $\bm A \, \bm\theta \leq \bm b$.
However, for the $V$-representation, we do not know of such a simple algebraic solution.
As a remedy, the following algorithm uses a geometric derivation to compute the conditional truncation boundaries of a parameter $\theta_{ij}$.

In Gibbs sampling, each step requires the distribution of the parameter $\theta_{ij}$ conditional on the current state of the remaining parameters $\bm\theta_{-ij}^{(t)}$.
Geometrically, this implies that, starting at the point $\bm\theta^{(t)}$,  we ``walk'' through the polytope in the direction of the $ij$-th dimension.
Since the polytope is convex, a straight line in this direction has two intersections with the convex hull of the vertices in $\bm V$ \citep[i.e., the lower and upper truncation boundaries $z_0$ and $z_1$, respectively;][]{lovasz1993random}.
It follows that the conditional truncation boundaries of the parameter $\theta_{ij}$ can be derived by computing these two intersections.

For this purpose, we solve two linear programs, one for the lower and one for the upper truncation boundary. 
By construction of the Gibbs sampler, the current sample $\bm \theta^{(t)}$ is known to be inside the polytope.
To formalize the intuition of ``walking'' in the direction of the $ij$-th dimension, we define the direction vector $\bm e^{(ij)}$ as the $ij$-th unit vector in $\mathbb R^D$ (with zero entries except for the $ij$-th entry, which equals one). 
The linear program now maximizes the distance $\lambda_1$ from the starting point $\bm\theta^{(t)}$ in the $ij$-th direction, under the constraint that the solution (i.e., the intersection) can be represented as a convex combination of the vertices $\bm v^{(s)}$:
\begin{align}
&\text{ maximize } \lambda_1 \,\,\,\,\, (\text{with } \lambda_1 \in \mathbb R \text{ and } \bm\alpha \in \mathbb R^S) \label{e.LP_intersection}\\
&\text{ subject to } \nonumber
\begin{cases}
\bm \theta^{(t)} + \lambda_1  \bm e^{(ij)} = \sum_{s=1}^S \alpha_s \bm v^{(s)}\\
\sum_{s=1}^S \alpha_s = 1 \\
\lambda_1>0 \text{ and } \alpha_s \geq 0 \text{ for all } s=1,\dots,S.
\end{cases}
\end{align}
The upper truncation boundary $z_1$ is then given by the $ij$-th coordinate of the intersection, $z_1 = \theta_{ij}^{(t)}+\lambda_1$.
The second intersection is computed by a linear program that maximizes the distance $\lambda_0$ in the opposite direction by replacing the left-hand side of the first constraint in Eq.~\eqref{e.LP_intersection} by $\bm \theta^{(t)} - \lambda_0 \, \bm e^{(ij)}$.
Accordingly, the lower truncation boundary is $z_0 = \theta_{ij}^{(t)} - \lambda_0$.

In summary, an iteration of the Gibbs sampler requires to solve the two linear programs in Eq.~\eqref{e.LP_intersection} and to use the resulting truncation boundaries for drawing a random sample from the scaled, truncated beta distribution in Eq.~\eqref{e.conditional_stbeta}.
As a starting value for the sampler, one can use a random, convex combination of the vertices $\bm v^{(s)}$.
However, to speed up convergence and reduce the number of burn-in samples (similarly as for the $Ab$-representation), we use the MAP estimate of the mixture weights $\hat{\bm\alpha}$ as a starting value. 
This is a valid strategy, even though the mixture weights $\bm\alpha$ are in general not identifiable because the corresponding estimate implied for the probability vector, $\hat{\bm\theta} = \sum_{s=1}^S \hat\alpha_s \bm v^{(s)}$, is still unique \citep{klauer2015parametric}.

\subsection{The Encompassing Bayes Factor}
\label{s.V_bf}

Similar as for the $Ab$-representation, Bayes factors for the $V$-representation can be computed with the encompassing method (cf. Section~\ref{s.Ab_bf}).
Essentially, this requires us to draw $T$ independent samples from a product-Dirichlet distribution (i.e., the prior and the posterior) and test for each sample $\bm\theta^{(t)}$ whether it is inside the convex hull of the vertices in the matrix $\bm V$.
A conceptually straightforward approach is to transform the $V$-representation to the $Ab$-representation \citep[e.g., using software such as \texttt{PORTA};][]{christof1997porta} and then check whether $\bm A \, \bm\theta \leq \bm b$ holds.
However, in a tutorial on polyhedral computation, \citet{fukuda2004there} refers to this approach as ``a method that we do not recommend but many people use.  
This method computes an inequality [$Ab$-] representation [...].
Once the system $\bm A \, \bm x \leq \bm b$ is computed, it is easy to check whether $\bm p$ [a specific vector to be tested] satisfies the system or not.
In most cases, this method is too expensive, since the convex hull computation is very hard in general and impossible for large data'' (meaning scenarios with a large number of vertices in the matrix $V$).  

As an alternative, \citet{fukuda2004there} recommends to directly  test whether a parameter vector $\bm\theta^{(t)}$ is inside the convex hull of the vertices $\bm v^{(s)}$ without computing the $Ab$-representation.
For this purpose, an algorithm is required to check whether there exists a set of nonnegative weights $\bm\alpha$ that sum to one and satisfy the constraint $\bm V^t \bm \alpha = \bm\theta^{(t)}$.
The computational approach is based on the geometric intuition that a vector is inside the polytope if and only if it is redundant for the definition of the convex hull. 
Based on this idea, \citet{fukuda2004there} shows that it is sufficient to solve the following linear program:
\begin{align*}
&\text{ maximize } \bm z^T \bm\theta^{(t)} - z_0 \,\,\,\,\, (\text{with } z_0 \in \mathbb R \text{ and } \bm z \in \mathbb R^D)\\
&\text{ subject to } 
\begin{cases}
\bm z^T \bm v^{(s)} - z_0 \leq 0 \text{ for all } s = 1, \dots, S\\
\bm z^T \bm\theta^{(t)} - z_0 \leq 1.
\end{cases}
\end{align*}
The parameter vector $\bm\theta^{(t)}$ is non-redundant (i.e., outside the convex hull) if and only if the optimal value of this linear program is strictly positive.
In this case, the solution $(z_0, \bm z)$ of the linear program implies that the vector $\bm\theta^{(t)}$ has a positive distance to the polytope (i.e., is outside of it) and thus is non-redundant for the convex hull. 

Despite the advantage of not having to compute the $Ab$-representation, this approach has the disadvantage that a linear program has to be solved for each prior or posterior sample $\bm\theta^{(t)}$.
For highly constrained models, this might require millions of samples, which renders the approach computationally costly. 
\citet{smeulders2018testing} addressed this challenge via a novel application of a column-generation algorithm -- however, this solution may not be general for arbitrary constraints. 
Future work may improve the efficiency of the algorithm by considering an extended formulation of the polytope \citep{davis-stober2017extended}.

\section{The R Package \texttt{multinomineq}}
\label{s.examples}

We implemented the above computational methods for multinomial models with convex, inequality constraints in C++ using the linear-algebra library Armadillo \citep{sanderson2010armadillo}. 
This has the advantage that many of the sequential computations can efficiently be performed using precompiled code.
To also make the methods available to a broad audience, the functions are embedded in the R package \texttt{multinomineq}, which is freely available on GitHub \citep[\url{www.github.com/danheck/multinomineq/};][]{heck2019multinomineq}.\footnote{
	The package will also be made available on CRAN.}
In the following, we show how to translate substantive theories into both the $Ab$- and the $V$-representation in R, how to estimate the inequality-constrained parameters $\bm\theta$, and how to test the constrained multinomial model by computing the encompassing Bayes factor.
R scripts for all analyses are available at the Open Science Framework (\href{https://osf.io/xv9u3/}{https://osf.io/xv9u3/}) and the package vignette provides more detailed explanations how to use the functions.%
    \footnote{ \url{https://www.dwheck.de/vignettes/multinomineq_intro.html}.}

\subsection{Introductory Example: Drug Dosage and Overconsumption}

In the drug dosage example in the introduction, \citet{paes1997impact} tested the hypothesis that overconsumption increases when the number of daily doses decreases.
Daily drug dosage was manipulated in a between-subjects design on three levels: once-daily ($n_1=40$), twice-daily, ($n_2=36$) and three times daily ($n_3=15$).
Across these conditions, the frequency of participants taking more tablets than prescribed was $k_1=16$, $k_2=4$, and $k_3=2$, respectively.
To test the substantive hypothesis of a monotonic relationship (i.e., $\theta_{11} \geq \theta_{21} \geq \theta_{31}$), one can use the inequality representation by defining a matrix $A$ and a vector $b$ in R:
\begin{lstlisting}
A <- matrix(c(-1,  1, 0,
               0, -1, 1),
            nrow = 2, byrow = TRUE)
b <- c(0, 0)
\end{lstlisting}
As defined in Eq.~\eqref{e.Ab}, the first row of \texttt{A} and the first value of the vector \texttt{b} define the inequality $-1 \cdot\theta_{11} + 1 \cdot\theta_{21} + 0 \cdot\theta_{31} \leq 0$.
Since we are working with binomial data, the function \texttt{bf\_binom} is used to compute the Bayes factor in favor of the order-constrained hypothesis:
\begin{lstlisting}
bf_binom(k = c(16, 4, 2), n = c(40, 36, 15), 
         A = A, b = b, M = 100000)
\end{lstlisting}
The vectors \texttt{k} and \texttt{n} provide the frequency of overconsumption and the number of observations per condition, respectively, whereas the argument \texttt{M} specifies how many prior and posterior samples are drawn from the encompassing model (the computation required 0.1 seconds on an Intel i7-7700).
By default, all functions in \texttt{multinomineq} assume a uniform prior on the restricted parameter space as specified via the shape parameters $\beta_{ij}=1$ of the Dirichlet distribution (this can be changed via the argument \texttt{prior}).

The function \texttt{bf\_binom} returns a matrix with the approximation of the Bayes factor in the first column and a summary of the sampling error in the remaining columns (i.e., the standard deviation and the 5\%- and 95\%-quantiles of the samples $B^{(r)}_{cu}$ from Section~\ref{s.bf_SE}):%
    \footnote{We opted for the label \texttt{se} to highlight that $\text{SD}(B^{(r)}_{cu})$ is conceptually similar to the standard error of the Bayes factor approximation.}
\begin{lstlisting}[basicstyle=\footnotesize\color{darkblack}\ttfamily,language=html]
         bf   se ci.5% ci.95%
bf_0u  2.11 0.02  2.08   2.14
bf_u0  0.47 0.00  0.47   0.48
bf_00' 2.70 0.03  2.65   2.75
\end{lstlisting}
Note that \texttt{bf\_0u = 2.11} refers to the Bayes factor for the constrained model versus the unconstrained model and \texttt{bf\_u0 = 0.47} to its inverse. 
In this example, the Bayes factor $B_{0u}=2.11$ indicates that the data provide only anecdotal evidence in favor of the hypothesis of monotonicity.
Moreover, the Bayes factor \texttt{bf\_00' = 2.70} compares the order-constrained model against an alternative model with a parameter space that is defined as the exact complement (i.e., $\Omega_{0'} = \Omega \setminus \Omega_0$).

\subsection{Testing Theories via the $V$-Representation: The Description-Experience Gap}
\label{s.ex_DEgap}

In the second example, we use the vertex representation to test the description-experience (DE) gap, which states that the presentation format of probabilities in risky gambles affects individuals' preferences \citep{hertwig2004decisions}.
In the \emph{description} condition, participants are presented with the exact numerical value of the probability of receiving a gain (e.\,g., ``you receive \$10 with $p=.20$ and \$0 otherwise'').
In the \emph{experience} condition, participants are presented with random samples of the two possible monetary outcomes \$10 and \$0 in sequential order, which occur with probabilities $p=.20$ and $p=.80$, respectively.
The DE gap states that rare events (i.\,e., small probabilities $p$) are overweighted in the description condition but underweighted in the experience condition.

\citet{hertwig2004decisions} tested the DE gap by presenting participants with six binary decision problems in each of the two experimental conditions.
These stimuli were constructed to ensure that over- and underweighting of small probabilities would result in distinct predicted patterns across the six decision problems.
However, under the assumption that preferences are heterogeneous across participants and trials, both over- and underweighting imply \emph{multiple} choice patterns that are in line with the psychological theory \citep{regenwetter2017constructbehavior}.
Given such heterogeneous predictions, observed choice frequencies can be modelled by a mixture distribution over the predicted patterns, and thus, by a multinomial model with inequality constraints (cf. Section~\ref{s.theory}).

To derive all patterns that are predicted when assuming heterogeneous preferences, \citet{regenwetter2017constructbehavior} repeatedly simulated preference patterns with cumulative prospective theory \citep{kahneman1979prospect} based on random data-generating parameters.
These patterns provide the vertex representation via the matrix $\bm V$ to define the inequality-constrained multinomial model.
For the DE gap, each row of the matrix $\bm V$ defines a specific pattern of predicted probabilities of choosing Option~H in each of the six decision problems.
The first of the 32 rows of the matrix \texttt{V} for the overweighting model are:
\begin{lstlisting}
       p1 p2 p3 p4 p5 p6 
 [1,]  0  0  0  0  0  0
 [2,]  0  0  0  0  0  1
 [3,]  0  0  0  1  0  0
 [4,]  ...
\end{lstlisting}

Bayesian parameter estimation for the parameters $\bm\theta$ proceeds by drawing posterior samples from the restricted model.
The corresponding Gibbs sampler for binomial data and the $V$-representation is called via:
\begin{lstlisting}
mcmc <- sampling_binom(k = c(9, 16, 16, 7, 12, 16), n = 25, 
                       V = V, M = 5000, cpu = 8)
\end{lstlisting}
Similar as in the first example, binomial data are defined via a vector \texttt{k} (frequencies of choosing Option~H for each decision problem) and a vector \texttt{n} (number of responses).
Here, the number of responses \texttt{n = 25} was identical for each lottery, and thus we can provide only a scalar value instead of a vector.
The Gibbs sampler for the $V$-representation required 62 seconds to sample \texttt{M = 5000} iterations on each of \texttt{cpu = 8} processing units in parallel.

This set of posterior samples is stored in the object \texttt{mcmc} containing a list of matrices, each with \texttt{M = 5000} posterior samples $\bm\theta^{(t)}$.
To assess the efficiency of the Gibbs sampler, we compared the number of MCMC iterations $M_\text{iter}=(5000-10)\cdot8$ (note that 10 samples are dropped as burnin) against the effective sample size $M_\text{eff}$, which is defined as the number of independent samples that would be required to achieve an equivalent estimation accuracy \citep{heck2018quantifying}. 
The corresponding ratio $M_\text{eff}/M_\text{iter}$ quantifies the loss in information due to dependent sampling and ranged between 0.80 and 1.00 for the six parameters (with an average of 0.94).
This shows that Gibbs sampling was very efficient for this specific example. 
Next, we can test model fit via posterior-predictive $p$-values (cf. Section~\ref{s.ppp}) which are computed as:
\begin{lstlisting}
ppp_binom(prob = mcmc, k = c(9, 16, 16, 7, 12, 16), n = 25)
\end{lstlisting}
For the data by \citet{hertwig2004decisions}, posterior-predictive $p$-values indicate that choices in the description condition were better described by over- than by underweighting ($p_B=.564$ vs. $p_B=.005$, respectively) whereas choices in the experience condition were better described by under- than by overweighting ($p_B=.587$ vs. $p_B=.058$, respectively).

Whereas Gibbs sampling is performed conditional on the inequality constraints, the encompassing Bayes factor requires prior and posterior samples from the \emph{unconstrained} model.  
Hence, the Bayes factor is implemented in a separate function: 
\begin{lstlisting}
bf <- bf_binom(k = c(9, 16, 16, 7, 12, 16), n = 25, 
               V = V, M = 5000, cpu = 8)
\end{lstlisting}
Using \texttt{M = 5000} samples on each of \texttt{cpu = 8} processing units required 34 seconds to compute the Bayes factor for the $V$-representation with $S=32$ vertices and $D=6$ parameters.
The data provided evidence for over- and against underweighting in the description condition ($B_{0u}=3.7$ versus $B_{u0} \approx 1,500$, respectively), while showing evidence for under- and against overweighting in the experience condition ($B_{0u}=34.8$ versus $B_{u0} \approx 1,100$, respectively; cf. Table~6 in \cite{regenwetter2017constructbehavior}).
Even though the substantive conclusions are similar to those of the original analysis in this specific example, there are examples in the literature where the reliance on multinomial models with inequality constraints can make a big difference (for instance, when testing transitivity of preferences; \cite{regenwetter2012behavioral}).  

The package \texttt{multinomineq} also provides a wrapper function \texttt{V\_to\_Ab} which transforms the vertex representation (i.e., the matrix $\bm V$) to the inequality representation by calling the R package \texttt{rPorta} \citep[][]{nunkesser2009rporta}.
If the transformation succeeds, one can rely on the computationally more efficient methods for the $Ab$-representation.
Moreover, the reverse transformation is available via the function \texttt{Ab\_to\_V}.

\subsection{Multinomial Data and Complex Inequalities: The Strict Weak Order Polytope}
\label{s.ex_transitivity}

In the last example, we test a set of complex inequality constraints using multinomial instead of binomial data.
We reanalyzed the data by \citet{regenwetter2012behavioral} who tested whether preferences for monetary gambles are transitive.
Participants were presented with 10 decision problems each featuring two out of the five lotteries $a,b,c,d,$ and $e$ (e.g., $a$ versus $b$; $a$ versus $c$; etc.).
A ternary choice format allowed for the opportunity to choose one of the two gambles or to respond ``indifferent.''
To test whether preferences are transitive under the assumption of heterogeneity, \citet{regenwetter2012behavioral} used an inequality-constrained multinomial model.
The parameter space of this model was defined by the strict weak order for the five lotteries which defines a polytope with 75,834 facet-defining inequalities for the 20 free parameters (i.\,e., the probabilities of choosing the first or the second gamble in each of the 10 problems).
In contrast to the previous example, these constraints are much more complex, as illustrated by the last three rows of the matrix $\bm A$:
\begin{lstlisting}[basicstyle=\scriptsize\color{darkblack}\ttfamily,language=html]
a>b b>a a>c c>a a>d d>a a>e e>a b>c c>b b>d d>b b>e e>b c>d d>c c>e e>c d>e e>d
  3   1  -1  -3   1   3  -3  -1   3   1  -3  -1  -1  -3  -1  -3   3   1   1   3
  3   1   1   3  -3  -1  -1  -3  -3  -1  -1  -3   3   1   1   3  -3  -1   1   3
  3   1   1   3  -1  -3  -3  -1  -3  -1   3   1  -1  -3  -3  -1   1   3   3   1
\end{lstlisting}
For illustration purposes, the R package \texttt{multinomineq} provides the data by \citet{regenwetter2012behavioral} in the data frame \texttt{regenwetter2012} and the inequality constraints of the strict weak order polytope for five alternatives in the list \texttt{swop5} (which contains the matrix $\bm A$ and the vector $\bm b$).

\begin{figure}[th]
\centering
\includegraphics[width=\linewidth]{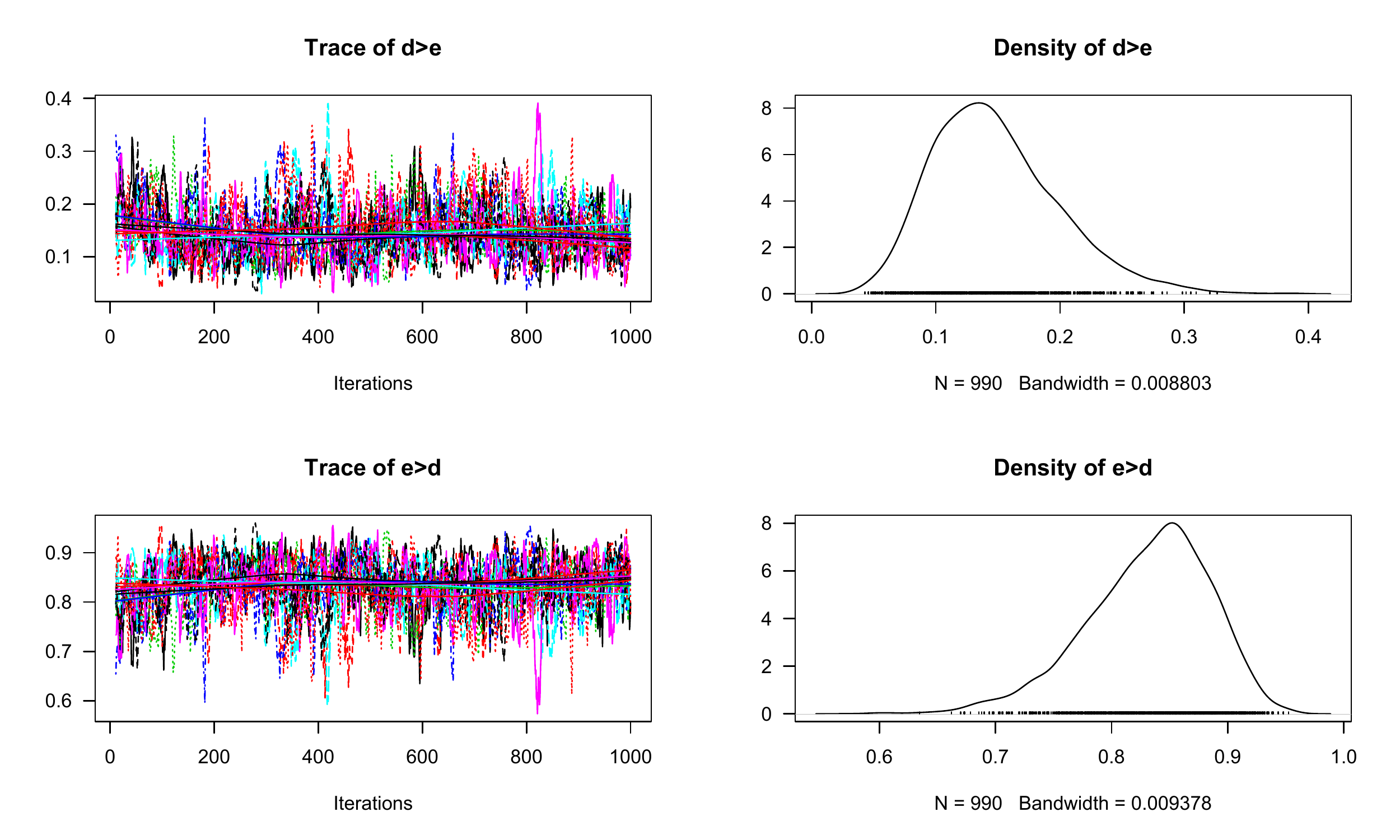}
\caption{
Posterior samples for two parameters of the 20-dimensional multinomial model with 75,834 inequality constraints to test a strict weak order for five choice alternatives \citep{regenwetter2012behavioral}.
}
\label{f.convergence}
\end{figure}

For multinomial response formats with more than two options per item type (e.g., ternary choice), the observed frequencies need to be provided in a different format than for binary responses.
In \texttt{multinomineq}, the vector \texttt{k} provides the observed frequencies of \emph{all} response options (ordered by item type: $k_{11},\dots,k_{1 J_1},k_{21},\dots,k_{2 J_2},\dots, k_{I,J_I}$) and the vector \texttt{options} provides the number of choice options per condition or item type (i.e., $J_1, J_2, \dots,J_I$).
In the present example, $I=10$ different item types with $J_i=3$ choice options were presented $n_i=45$ times each.
To fit the data of the first participant of \citet{regenwetter2012behavioral},  posterior samples from the inequality-constrained model are drawn via:
\begin{lstlisting}
mcmc <- sampling_multinom(k = c(21,24,0,  2,43,0,  0,45,0,  ...), 
                          options = c(3,3,3,3,3,3,3,3,3,3), 
                          A = A, b = b, M = 1000, cpu = 8)
\end{lstlisting}
The Gibbs sampler for the $Ab$-representation with 75,834 facet-defining inequalities on 20 parameters required approximately two minutes to sample \texttt{M = 1000} iterations on each of \texttt{cpu = 8} processing units in parallel.
Figure~\ref{f.convergence} shows the posterior samples for two probability parameters, indicating that the Gibbs sampler converged very fast (only 10 samples were discarded as burn-in) and resulted in acceptable autocorrelations of 0.73, 0.36, and 0.19 for lags of 1, 5, and 10, respectively (averaged across parameters).
Similarly as in the previous example, the ratio $M_\text{eff}/M_\text{iter}$ quantifies the loss in information and ranged between 0.04 and 0.31 for the 20 parameters (with an average of 0.14).
This shows that the efficiency of the implemented Gibbs sampling is acceptable even for complex inequality constraints.  

To compute the encompassing Bayes factor for the $Ab$-representation, the function \texttt{bf\_multinom} can be applied similarly as shown in the previous example.
Here, we compute the Bayes factor more efficiently by using separate calls to the function \texttt{count\_multinom} to count the number of samples that are inside the inequality-constrained parameter space $\Omega_c$ for (a) the prior distribution and (b) the posterior distribution. 
This is advantageous since the proportion $\hat c$ of prior samples satisfying the constraints is required repeatedly for the analysis of all participants.
To increase precision, we used a larger number of samples to approximate the normalizing constant $c$ (which required 129 seconds):
\begin{lstlisting}
prior <- count_multinom(k = 0, options = rep(3, 10),
                        A = A, b = b, M = 1e+7, cpu = 8)
\end{lstlisting}
To count posterior samples, we applied the automatic procedure by defining a vector \texttt{steps} that splits the inequalities of the $Ab$-representation into nested subsets.
For instance, the argument \texttt{steps = c(5000,10000,15000,...)} will generate a sequence of intermediate nested models that include only the first 5,000 (then the first 10,000; the first 15,000 etc.) inequalities.
Moreover, the integer \texttt{cmin} defines the minimum number of samples $T_\text{min}$ per step that must be inside the constrained parameter space $\Omega_c$ (cf. Section~\ref{s.automatic}).
Here, we use \texttt{steps = 75834} to specify that only one step should be used (i.e., from the encompassing model to the restricted model with all inequalities):
\begin{lstlisting}
post <- count_multinom(k = c(21,24,0,  2,43,0,  0,45,0, ...), 
                       options = rep(3, 10), A = A, b = b, cpu = 8,
                       M = 10000, steps = 75834, cmin = 10)
\end{lstlisting}
Here, \texttt{M = 10000} posterior samples were repeatedly drawn until the minimum number of samples \texttt{cmin = 10} inside the parameter space was reached (which required 39 seconds).
Finally, the function \texttt{count\_to\_bf} combines the prior and posterior proportions into the encompassing Bayes factor and uses beta distributions to approximate estimation uncertainty (cf. Section~\ref{s.bf_SE}):
\begin{lstlisting}
count_to_bf(prior, post)
\end{lstlisting}

In their paper, \citet{regenwetter2012behavioral} reported whether the observed choice proportions of 30 participants across three gamble sets satisfied the inequality constraints descriptively.
Moreover, frequentist $p$-values were used to test whether the discrepancy of the observed choice frequencies to the inequality-constrained model was significant \citep{davis-stober2009analysis}.
To complement the original analysis, Table~\ref{t.regenwetter2012} provides the corresponding Bayes factors for these data with a uniform prior on the choice probabilities (the complete analysis of the 90 data sets required 54 minutes).  
The first two columns of Table~\ref{t.regenwetter2012} were previously reported by \citet{davis-stober2015individual}, who calculated the Bayes factors using the encompassing prior approach with rejection sampling.  
There are some minor discrepancies between the values reported in our Table~\ref{t.regenwetter2012} and \citet{davis-stober2015individual}'s values, reported in their Tables~4 and 5.  
These discrepancies were due to an error in the original computer code used by \citet{davis-stober2015individual} that under-estimated the volume of the strict weak order polytope. 
Upon correction, both sets of analyses agree, see \citet{davis-stober2018errata} for an erratum.

In Table~\ref{t.regenwetter2012}, Bayes factors for most participants and gamble sets are clearly larger than one, thus providing evidence that preferences are transitive.
Moreover, the Bayes factor $B_{0u}$ takes into account how well the inequality-constrained model describes the data:
If the inequality constraints are only barely met (i.e., if the observed choice proportions are close to the facets of the polytope), the Bayes factor $B_{0u}$ will be smaller compared to the case that the constraints are clearly satisfied (i.e., if the distance is large between the observed choice proportions and the facets of the polytope).
In that latter case, the proportion of posterior samples inside the polytope will be larger compared to the former case.
In Table~\ref{t.regenwetter2012}, this is reflected by the substantial variance in how close the Bayes factors are to the maximum Bayes factor $B^\text{max}_{0u} = 1/c \approx 2187$ (which is obtained if all posterior samples satisfy the constraints).

\begin{table}[H]
\footnotesize
\begin{threeparttable}
\caption{Bayes factors for the ternary choice data by \citet{regenwetter2012behavioral}.}
\label{t.regenwetter2012}
\begin{tabular}{rrrrrrr}
  \hline
ID & Set I: $B_{0u}$ & (SD) & Set II: $B_{0u}$ & (SD) & Set III: $B_{0u}$ & (SD) \\ 
  \hline
    1 & 1680.61 & 9.48 & 2079.92 & 11.05 & 502.91 & 4.18 \\ 
    2 & 157.51 & 2.14 & 2081.91 & 10.88 & 135.93 & 1.99 \\ 
    3 & 1667.22 & 9.29 & 387.64 & 3.48 & 141.94 & 2.06 \\ 
    4 & 563.26 & 4.56 & 2116.39 & 11.09 & 0.08 & 0.01 \\ 
    5 & 636.76 & 4.84 & 297.48 & 3.07 & 652.85 & 4.96 \\ 
    6 & 820.96 & 5.63 & 1367.04 & 8.15 & 1802.25 & 9.75 \\ 
    7 & 2138.00 & 11.22 & 2030.90 & 10.87 & 2085.27 & 10.81 \\ 
    8 & 910.60 & 6.11 & 1776.32 & 9.55 & 1667.60 & 9.30 \\ 
    9 & 72.51 & 1.44 & 0.33 & 0.04 & 2010.30 & 10.74 \\ 
   10 & 269.69 & 2.85 & 1623.13 & 9.12 & 1225.92 & 7.55 \\ 
   11 & 117.57 & 1.87 & 180.71 & 2.30 & 246.66 & 2.82 \\ 
   12 & 1266.41 & 7.56 & 910.60 & 6.00 & 1486.30 & 8.54 \\ 
   13 & 41.61 & 1.09 & 42.27 & 1.07 & 1644.19 & 9.06 \\ 
   14 & 1.86 & 0.18 & 64.37 & 1.37 & <0.01 & <0.01 \\ 
   15 & 117.29 & 1.86 & 280.76 & 2.94 & 107.48 & 1.76 \\ 
   16 & 1996.26 & 10.39 & 2097.10 & 11.25 & 1000.49 & 6.63 \\ 
   17 & 964.54 & 6.44 & 979.67 & 6.53 & 692.88 & 5.20 \\ 
   18 & 2132.92 & 11.15 & 1106.72 & 7.06 & 0.54 & 0.05 \\ 
   19 & 2159.56 & 11.16 & 1289.15 & 7.64 & 2099.53 & 11.19 \\ 
   20 & 1630.53 & 9.15 & 1336.71 & 7.90 & 2063.55 & 10.90 \\ 
   21 & 1757.98 & 9.62 & 1322.86 & 7.81 & 143.06 & 2.06 \\ 
   22 & 0.87 & 0.09 & 0.01 & <0.01 & 207.84 & 2.45 \\ 
   23 & 720.53 & 5.30 & 845.69 & 5.84 & 146.50 & 2.08 \\ 
   24 & 701.07 & 5.09 & 928.01 & 6.34 & 1601.10 & 9.01 \\ 
   25 & 717.00 & 5.25 & 742.38 & 5.25 & 23.09 & 0.81 \\ 
   26 & 39.32 & 1.05 & 49.21 & 1.20 & 489.69 & 4.15 \\ 
   27 & 1086.04 & 6.97 & 955.14 & 6.25 & 216.01 & 2.61 \\ 
   28 & 1553.07 & 8.72 & 1852.24 & 10.02 & 437.80 & 3.85 \\ 
   29 & 0.41 & 0.04 & <0.01 & <0.01 & 1152.67 & 7.15 \\ 
   30 & 54.53 & 1.24 & 270.27 & 2.86 & 1152.73 & 7.14 \\ 
   \hline
\end{tabular}
\vspace{.1cm}
\begin{tablenotes}
\footnotesize
\textit{Note.} 
The Bayes factor $B_{0u}$ quantifies the evidence for the inequality-constrained multinomial model $\mathcal M_c$, which assumes transitive preferences (i.e., a strict weak order), against the encompassing, unconstrained model $\mathcal M_u$.  
The posterior standard deviation (SD) of the Bayes factor approximation provides the numerical error due to Monte Carlo sampling.
\end{tablenotes}
\end{threeparttable}
\end{table}

\section{Discussion}
\label{s.discussion}

In mathematical psychology in general and judgment and decision making in particular, many theories can be formulated by a set of linear inequality constraints on multinomial models \citep{iverson2006essay}.
This includes representational measurement theory \citep{krantz1971foundations, karabatsos2001rasch}, state-trace analysis \citep{prince2012design}, decision axioms such as transitivity \citep{regenwetter2011transitivity, myung2005bayesian}, random utility models \citep[for a review, see][]{marley2017choice}, and multiattribute probabilistic inference \citep{heck2017information}.
Moreover, inequality constraints are often of interest in the analysis of contingency tables \citep{lindley1964bayesian, klugkist2010bayesian} and in cognitive diagnostic assessment \citep{hoijtink2014cognitive}. 
Geometrically, linear order constraints on the probabilities $\bm\theta$ result in a parameter space representing a convex polytope, which can be defined either by a set of inequalities (i.\,e., the $Ab$-representation: $\bm A \, \bm \theta \leq \bm b$) or by the convex hull of a set of predicted patterns $\bm v^{(s)}$ for the choice frequencies (i.\,e., the $V$-representation: $\bm\theta = \sum_s \alpha_s \bm v^{(s)}$ with nonnegative weights $\bm\alpha$).
To facilitate statistical tests of such models using either type of representation, we generalized existing methods for Bayesian inference.
Specifically, we developed a Gibbs sampler to draw posterior samples and improved on the encompassing Bayes factor approach. 
An efficient C++ implementation of the methods is provided in the R package \texttt{multinomineq} \citep{heck2019multinomineq}.

In practical applications, the following three steps are required to test a multinomial model with linear inequality constraints. 
First, it is necessary to define the restricted parameter space.
For some substantive theories, it might be possible to derive a set of inequality constraints directly \citep[e.\,g.,][]{koppen1995random, hilbig2014generalized}.
Otherwise, it is necessary to derive all deterministic response patterns $\bm v^{(s)}$ that are predicted by a substantive theory \citep{regenwetter2017constructbehavior}.
This provides the vertices defining a convex hull for the $V$-representation. 
To increase the efficiency of the sampling methods for Bayesian inference, this $V$-representation can be converted into the inequality ($Ab$-) representation using convex hull algorithms \citep{christof1997porta,assarf2017computing}.
Second, a Gibbs sampler allows drawing random samples from the posterior distribution of the probability parameters $\bm\theta$ conditional on the inequality constraints that are defined by the matrix $\bm V$ or by the matrix $\bm A$ and the vector $\bm b$.
These samples, in turn, allow testing the fit of the model by posterior-predictive $p$-values \citep{meng1994posterior}, an approach conceptually similar to a parametric bootstrap for Pearson's $X^2$.
Third, to compare several competing models, the Bayes factor quantifies the relative evidence for the constrained versus the unconstrained, encompassing model.
The Bayes factor has the advantage that it takes model complexity into account which is defined as the relative volume of the constrained parameter space when assuming a uniform prior for the parameters \citep{hoijtink2011informative}.
Accordingly, models with large (or unconstrained) parameter spaces are penalized more for complexity than parsimonious models with small parameter spaces.
Since the approximation of the Bayes factor is usually computationally expensive (or even infeasible), one may restrict model selection to those models that have an adequate fit in terms of posterior-predictive $p$-values \citep{regenwetter2018heterogeneity}.

\subsection{Analysis of Nested Data Structures} 
\label{s.aggregation}

Many psychological experiments use repeated measures per person, and thus, the natural question arises whether and how inequality-constrained multinomial models can account for nested data structures.
As a first strategy, it is possible to fit a single inequality-constrained model to the summed individual frequencies (often referred to as ``complete pooling'').
Even if there is heterogeneity in the parameter vector $\bm\theta$ across participants, this is a valid approach because the mean of the individual parameters must satisfy the inequality constraints if they hold for each person on the individual level.
This is due to the fact that (between-person) mixtures of (within-person) mixture distributions do not affect the convex, inequality-constrained parameter space \citep{regenwetter2018role}.
However, the reverse statement does not hold: it is possible that the inequality constraints are satisfied by the aggregated frequencies even though they are violated by some or all of the participants.  

To detect variability across participants, researchers often test inequality-constrained models for each person separately, for instance, using Bayesian model selection on the individual level \citep[e.\,g.,][]{heck2017information, regenwetter2014qtest}.
This ``no pooling''-strategy leads to one Bayes factor per person, which opens the question of how to aggregate the distribution of individual Bayes factors on the population level.
If a researcher wants to test whether the restricted model accounts for the data by \emph{all} participants, one can compute the group Bayes factor defined as the product of the individual Bayes factors \citep{klaassen2018all}.
Often, however, researchers are interested in classifying participants as users of different decision strategies which are operationalized by separate models \citep{heck2017information} and  want to know whether a \emph{majority} of participants are better described by one model than by the competitors.

To test hypotheses about latent classes of participants, \citet{cavagnaro2018modelbased} proposed a random-effects model that only requires on one or more distributions of individual Bayes factors (e.g., for a control and treatment condition).
The random-effects Bayes factor assumes that participants are clustered into latent classes that differ with respect to the data-generating model.
The method then tests whether the group sizes of these latent classes differ within or across conditions \citep[for details, see][]{cavagnaro2018modelbased}.
The technical framework and implementation of this approach is closely related to the multinomial models discussed in the present manuscript, with the main difference that the individual Bayes factors and not the raw choice frequencies serve as input for statistical inference.%
    \footnote{In the random-effects model, the probability parameters $\theta_{ij}$ are the latent class probabilities for model $j$ in condition $i$.
	Using variational Bayesian inference \citep{stephan2009bayesian}, the observed frequencies $k_{ij}$ are simply replaced by the sum of the individual posterior model probabilities across individuals, $k'_{ij} = \sum_m p(y^{(m)}_i  \mid \mathcal M_j)$, where $y_i^{(m)}$ are  the raw data of  person $m$ in the $i$-th condition.
	Posterior model probabilities are computed via $p(y^{(m)}_i  \mid \mathcal M_j) =  B_{j0}^{(m)} \large/ \sum_k B_{k0}^{(m)}$ when assuming equal prior odds for the models.	}
Hence, the methods developed in the present manuscript allow computing the random-effects population Bayes factor to test, for instance, whether a specific treatment results in a larger proportion of risk-seeking versus risk-avoiding participants \citep{cavagnaro2018modelbased}.
Overall, inequality-constrained multinomial models (and their implementation in the R package \texttt{multinomineq}) may thus serve as a general and comprehensive framework for testing theories both on the individual and the population level.

As a third alternative (called ``partial pooling''), it is possible to develop hierarchical multinomial models with inequality constraints to account for nested data structures.
For example, one could assume that the inequality constraints hold for all participants but that the individual parameter vectors $\bm\theta$ differ across participants \citep{haaf2019some}.
To model this substantive hypothesis, one may define a separate parameter vector $\bm\theta^{(p)}_{ij}$ for each participant $p$ and assume a Dirichlet distribution of $\bm\theta^{(p)}$ on the group level.
Such a hierarchical approach offers the benefit of shrinkage, meaning that the individual parameter estimates will be informed by each other and pulled towards the group mean \citep{efron1975data}.
However, future work is required to develop computational methods fur such high-dimensional models. 
There are also unresolved questions regarding the direction of parameter shrinkage and how this could interact with the substantive models being tested.

\subsection{Limitations and Future Directions}
\label{s.future}

Several open questions and possible limitations remain concerning computational methods for inequality-constrained multinomial models.
First, simulation studies are required to assess the efficiency of the developed methods and to assess the benefit of considering the structure of a given set of constraints. 
For instance, if the parameter space is highly regular such as that of a linear order polytope \citep{regenwetter2011transitivity}, analytical or computational improvements might be possible to speed up estimation and testing of a model \citep{davis-stober2017extended}.
Moreover, the algorithms could be adapted by taking into account which of the inequality constraints are violated for a given dataset.
For instance, if 95\% of the constraints are satisfied descriptively, it is likely that most of the posterior samples from the encompassing model will also adhere to these constraints (see \cite{smeulders2018testing} for applications of similar computational heuristics).
To approximate the Bayes factor, it might thus be beneficial to reorder the inequalities in the $Ab$-representation by their relative strength (i.e., by the rejection probability).
Similarly, in the stepwise and automatic procedures, the relative strength of the inequalities could be exploited to establish a more efficient clustering into nested models.

Substantively, future work should investigate the choice of prior distribution for multinomial models with inequality constraints.
When relying on Bayesian model selection, priors should reflect the psychological theory underlying a specific model and domain as closely as possible \citep{lee2018determining}.
The computational methods outlined in the present paper assume independent Dirichlet distributions for the choice probabilities, which includes the uniform distribution as a special case \citep[e.g.,][]{hoijtink2014cognitive}.%
    \footnote{The uniform prior does not belong to the class of objective priors, which are derived based on asymptotic considerations \citep{ghosh2011objective}.}
In the absence of further knowledge, a uniform prior is often justified since researchers want to assign equal probability to all parameters that are in line with a substantive theory.
Moreover, using simulations, \citet{klugkist2010bayesian} showed that a uniform prior results in a good performance of the encompassing Bayes factor for inequality-constrained multinomial models.
The assumption of independence may be more controversial, especially if the same choice alternatives are presented repeatedly across different paired comparisons \citep{regenwetter2018role}.
As an alternative to independent Dirichlet priors, \citet{mccausland2013prior} derived a family of prior distributions for choice probabilities on non-empty subsets of a finite set of objects that take dependency into account.
However, this prior is limited to very specific types of decision-making models and not suited as a default for the general class of inequality-constrained multinomial models presented in the present paper.
As a remedy, the independent Dirichlet prior may serve as a useful approximation for testing psychological theories and as a convenient default for inequality-constrained data analysis in general.
This view is in line with the existing literature as indicated by the conclusion of \citet[][p.~6]{regenwetter2018role} that ``statistical analyses of these models usually depend on auxiliary independence and stationarity assumptions to get simple and tractable test statistics''.

\begin{figure}[th]
\centering
\includegraphics[width=\linewidth]{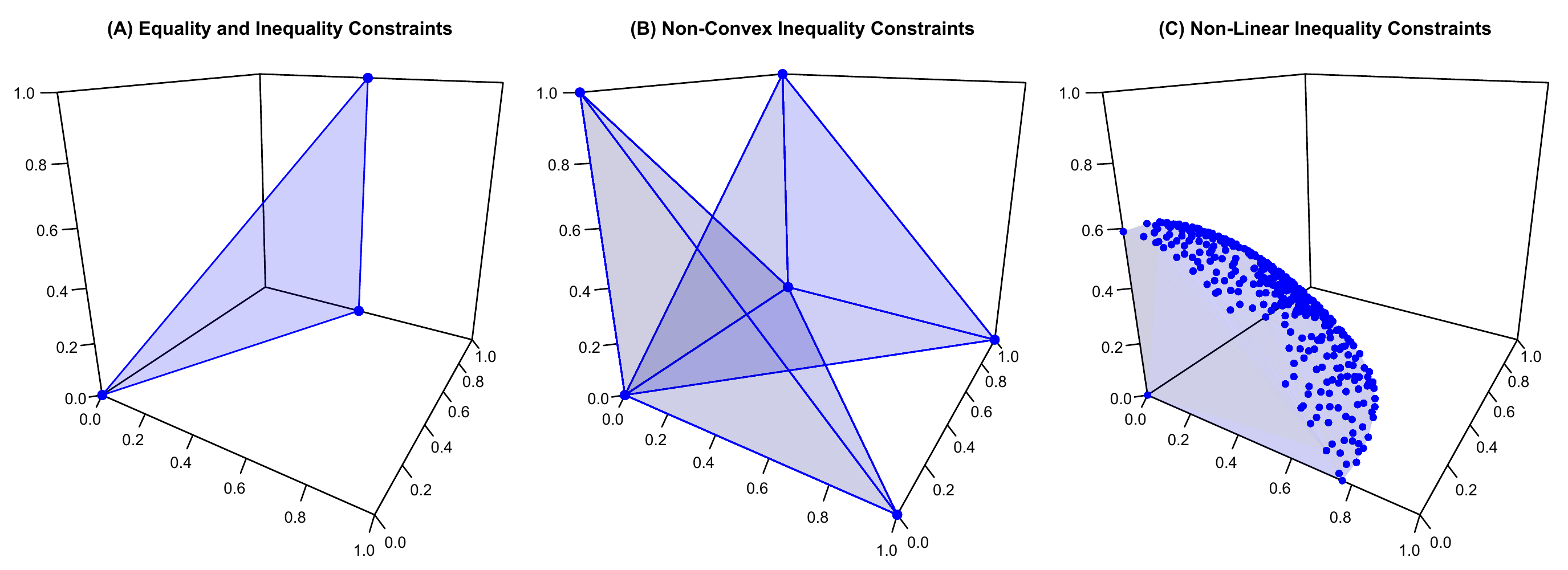}
\caption{
The methods developed in the present paper apply to multinomial models with convex linear inequality constraints (cf. Figure~\ref{f.3d}).
Future computational developments may focus on (A) models with both equality \emph{and} convex inequality constraints that result in a lower-dimensional parameter space, (B) models with non-convex inequality constraints such as (partially overlapping) unions of two or more convex polytopes, (C) models with nonlinear inequalities.
}
\label{f.future}
\end{figure}

Concerning the type of models that can be analyzed, the methods developed in the present manuscript only apply to convex, linear inequality constraints that result in a parameter space of full dimensionality.
For instance, a model for $D=3$ binomial probabilities must have three free parameters, as represented by a 3-dimensional polytope (cf. Figure~\ref{f.3d}).
However, theories often predict that choice probabilities are identical across different item types, which leads to \emph{equality} constraints of the form $\theta_{ij}=\theta_{kl}$ \citep[e.\,g.,][]{broder2003bayesian}.
Formally, any set of linear equality constraints can be defined via a matrix $\bm C$ and a vector $\bm d$ similar to the $Ab$-representation of inequalities (i.\,e., by $\bm C \,\bm\theta = \bm d$), thereby defining in a lower-dimensional parameter space.
For instance, in Figure~\ref{f.future}A, the 3-dimensional parameter space $\Omega$ reduces to a constrained parameter space represented by a 2-dimensional plane.
For models posing both equality and inequality constraints, the Gibbs sampler in Section~\ref{s.Ab_gibbs} needs to be adapted to walk through the lower-dimensional projection of the $D$-dimensional space of choice probabilities.
Moreover, the encompassing approach of computing Bayes factors will fail for such models because random samples from the $D$-dimensional prior or posterior distribution will be outside the lower-dimensional parameter space with probability one.
As a remedy, the encompassing Bayes factor method allows to test ``about equality constraints'' defined as $\mid\bm C \,\bm\theta - \bm d \mid \leq \delta$ which provide similar results as $\delta \rightarrow 0$ \citep{hoijtink2011informative, klugkist2010bayesian}.

Other generalizations of the presented methods concern different types of inequality-constraints.
Models with \emph{non-convex} parameter spaces may emerge if a theory combines multiple inequality constraints by logical OR statements as shown in the error model in Figure~\ref{f.pred}B. 
Whereas the conjunction of multiple inequalities (e.\,g., $\theta_{11}\leq \theta_{21}$ AND $\theta_{11}\leq \theta_{31}$ etc.) always results in a convex parameter space, a disjunction (e.\,g., $\theta_{11} + \theta_{21} +\theta_{31}\leq 1$ OR $\theta_{11} + \theta_{31} \leq \theta_{21}$) may result in a non-convex parameter space.
The latter case is illustrated in Figure~\ref{f.future}B, showing that the parameter space of a disjunction is a (possibly overlapping) union of convex polytopes.
To analyze such models, the encompassing Bayes factor can again be approximated by counting how many prior and posterior samples fall into the non-convex parameter space.
However, the Gibbs sampler developed in Section~\ref{s.Ab_gibbs} requires a convex parameter space to ensure that the support of the conditional posterior distribution is always a connected interval with one lower and one upper truncation boundary, which is not necessarily the case for non-convex parameter spaces.

An important type of restrictions are \emph{nonlinear} inequality constraints as those shown in Figure~\ref{f.future}C. 
Nonlinear inequalities emerge in the analysis of contingency tables \citep{klugkist2010bayesian} and can be represented neither by a finite number of vertices nor by linear inequalities. 
Instead, a general indicator function $\mathbb I_{\Omega_n}(\bm\theta)$ is required to define the restricted parameter space $\Omega_n \subset \Omega$.
Again, the encompassing Bayes factor can be computed by counting the number of prior and posterior samples for which $\mathbb I_{\Omega_n}(\bm\theta)= 1$.
It is also possible to adapt the Gibbs sampler in Section~\ref{s.Ab_gibbs} to sample from the posterior distribution as long as the parameter space is convex.
In this case, only the computation of the lower and upper truncation boundaries of the conditional posterior distribution (i.e., the scaled, truncated beta in Eq.~\eqref{e.conditional_stbeta}) needs to be generalized for nonlinear indicator functions (e.g., via a bisection algorithm).
The function \texttt{sampling\_nonlinear} provides a Gibbs sampler for nonlinear inequality constraints defined via an indicator function $\mathbb I_{\Omega_n}(\bm\theta)$.

\subsection{Conclusion}

To test psychological theories, it is important that statistical models reflect a theory's core predictions without requiring strong and often arbitrary auxiliary assumptions.
Multinomial models with inequality constraints provide an ideal framework for this purpose and allow to test both formal theories assuming deterministic axioms \citep{iverson2006essay} as well as verbal theories predicting multiple choice patterns \citep{regenwetter2017constructbehavior}.
Given the implementation of Bayesian inference for this model class in the R package \texttt{multinomineq} \citep{heck2019multinomineq}, it will thus become easier for researchers to test psychological theories. 

\small
\bibliography{bibtex}

\ifdefined\DeclarePrefChars\DeclarePrefChars{'’-}\else\fi
\begin{thebibliography}{}

\bibitem[Agresti and Hitchcock, 2005]{agresti2005bayesian}
Agresti, A. and Hitchcock, D.~B. (2005).
\newblock Bayesian inference for categorical data analysis.
\newblock {\em Statistical Methods and Applications}, 14:297--330, DOI:
  \href{https://dx.doi.org/10.1007/s10260-005-0121-y}{\ttfamily
  10.1007/s10260-005-0121-y}.

\bibitem[Assarf et~al., 2017]{assarf2017computing}
Assarf, B., Gawrilow, E., Herr, K., Joswig, M., Lorenz, B., Paffenholz, A., and
  Rehn, T. (2017).
\newblock Computing convex hulls and counting integer points with polymake.
\newblock {\em Mathematical Programming Computation}, 9:1--38, DOI:
  \href{https://dx.doi.org/10.1007/s12532-016-0104-z}{\ttfamily
  10.1007/s12532-016-0104-z}.

\bibitem[Avis et~al., 1997]{avis1997how}
Avis, D., Bremner, D., and Seidel, R. (1997).
\newblock How good are convex hull algorithms?
\newblock {\em Computational Geometry}, 7:265--301, DOI:
  \href{https://dx.doi.org/10.1016/S0925-7721(96)00023-5}{\ttfamily
  10.1016/S0925-7721(96)00023-5}.

\bibitem[Bamber and van Santen, 2000]{bamber2000how}
Bamber, D. and van Santen, J. P.~H. (2000).
\newblock How to assess a model's testability and identifiability.
\newblock {\em Journal of Mathematical Psychology}, 44:20--40, DOI:
  \href{https://dx.doi.org/10.1006/jmps.1999.1275}{\ttfamily
  10.1006/jmps.1999.1275}.

\bibitem[Barlow et~al., 1972]{barlow1972statistical}
Barlow, R.~E., Bartholomew, D.~J., Bremner, J.~M., and Brunk, H.~D. (1972).
\newblock {\em Statistical inference under order restrictions: Theory and
  application of isotonic regression}.
\newblock {Wiley}, {London, New York}.

\bibitem[Br{\"{o}}der and Schiffer, 2003]{broder2003bayesian}
Br{\"{o}}der, A. and Schiffer, S. (2003).
\newblock Bayesian strategy assessment in multi-attribute decision making.
\newblock {\em Journal of Behavioral Decision Making}, 16:193--213, DOI:
  \href{https://dx.doi.org/10.1002/bdm.442}{\ttfamily 10.1002/bdm.442}.

\bibitem[Br{\o}ndsted, 2012]{brondsted2012introduction}
Br{\o}ndsted, A. (2012).
\newblock {\em An introduction to convex polytopes}.
\newblock {Springer}, {New York, NY}.

\bibitem[Carbone and Hey, 2000]{carbone2000which}
Carbone, E. and Hey, J.~D. (2000).
\newblock Which error story is best?
\newblock {\em Journal of Risk and Uncertainty}, 20:161--176, DOI:
  \href{https://dx.doi.org/10.1023/A:1007829024107}{\ttfamily
  10.1023/A:1007829024107}.

\bibitem[Cavagnaro and Davis-Stober, 2018]{cavagnaro2018modelbased}
Cavagnaro, D.~R. and Davis-Stober, C.~P. (2018).
\newblock A model-based test for treatment effects with probabilistic
  classifications.
\newblock {\em Psychological Methods}, 23:672--689, DOI:
  \href{https://dx.doi.org/10.1037/met0000173}{\ttfamily 10.1037/met0000173}.

\bibitem[Christof et~al., 1997]{christof1997porta}
Christof, T., L{\"{o}}bel, A., and Stoer, M. (1997).
\newblock {{PORTA}} - polyhedron representation transformation algorithm.
\newblock \url{http://porta.zib.de/}.

\bibitem[Cyrus and Beck, 1978]{cyrus1978generalized}
Cyrus, M. and Beck, J. (1978).
\newblock Generalized two- and three-dimensional clipping.
\newblock {\em Computers \& Graphics}, 3:23--28, DOI:
  \href{https://dx.doi.org/10.1016/0097-8493(78)90021-3}{\ttfamily
  10.1016/0097-8493(78)90021-3}.

\bibitem[Davis-Stober, 2009]{davis-stober2009analysis}
Davis-Stober, C.~P. (2009).
\newblock Analysis of multinomial models under inequality constraints:
  {{Applications}} to measurement theory.
\newblock {\em Journal of Mathematical Psychology}, 53:1--13, DOI:
  \href{https://dx.doi.org/10.1016/j.jmp.2008.08.003}{\ttfamily
  10.1016/j.jmp.2008.08.003}.

\bibitem[Davis-Stober, 2012]{davis-stober2012lexicographic}
Davis-Stober, C.~P. (2012).
\newblock A lexicographic semiorder polytope and probabilistic representations
  of choice.
\newblock {\em Journal of Mathematical Psychology}, 56:86--94, DOI:
  \href{https://dx.doi.org/10.1016/j.jmp.2012.01.003}{\ttfamily
  10.1016/j.jmp.2012.01.003}.

\bibitem[Davis-Stober et~al., 2015]{davis-stober2015individual}
Davis-Stober, C.~P., Brown, N., and Cavagnaro, D.~R. (2015).
\newblock Individual differences in the algebraic structure of preferences.
\newblock {\em Journal of Mathematical Psychology}, 66:70--82, DOI:
  \href{https://dx.doi.org/10.1016/j.jmp.2014.12.003}{\ttfamily
  10.1016/j.jmp.2014.12.003}.

\bibitem[Davis-Stober et~al., 2018a]{davis-stober2018errata}
Davis-Stober, C.~P., Brown, N., and Cavagnaro, D.~R. (2018a).
\newblock Erratum to {{Davis}}-{{Stober}} et al. (2015): {{Individual}}
  differences in the algebraic structure of preference.

\bibitem[Davis-Stober et~al., 2018b]{davis-stober2017extended}
Davis-Stober, C.~P., Doignon, J.-P., Fiorini, S., Glineur, F., and Regenwetter,
  M. (2018b).
\newblock Extended formulations for order polytopes through network flows.
\newblock {\em Journal of Mathematical Psychology}, 87:1--10, DOI:
  \href{https://dx.doi.org/10.1016/j.jmp.2018.08.003}{\ttfamily
  10.1016/j.jmp.2018.08.003}.

\bibitem[Davis-Stober et~al., 2016]{davis-stober2016bayes}
Davis-Stober, C.~P., Morey, R.~D., Gretton, M., and Heathcote, A. (2016).
\newblock Bayes factors for state-trace analysis.
\newblock {\em Journal of Mathematical Psychology}, 72:116--129, DOI:
  \href{https://dx.doi.org/10.1016/j.jmp.2015.08.004}{\ttfamily
  10.1016/j.jmp.2015.08.004}.

\bibitem[Devroye, 1986]{devroye1986nonuniform}
Devroye, L. (1986).
\newblock {\em Non-Uniform Random Variate Generation}.
\newblock {Springer}, {New York, NY}.

\bibitem[Doignon and Rexhep, 2016]{doignon2016primary}
Doignon, J.-P. and Rexhep, S. (2016).
\newblock Primary facets of order polytopes.
\newblock {\em Journal of Mathematical Psychology}, 75:231--245, DOI:
  \href{https://dx.doi.org/10.1016/j.jmp.2016.07.004}{\ttfamily
  10.1016/j.jmp.2016.07.004}.

\bibitem[Efron and Morris, 1975]{efron1975data}
Efron, B. and Morris, C. (1975).
\newblock Data analysis using {{Stein}}'s estimator and its generalizations.
\newblock {\em Journal of the American Statistical Association}, 70:311--319,
  DOI: \href{https://dx.doi.org/10.1080/01621459.1975.10479864}{\ttfamily
  10.1080/01621459.1975.10479864}.

\bibitem[Fishburn, 1992]{fishburn1992induced}
Fishburn, P.~C. (1992).
\newblock Induced binary probabilities and the linear ordering polytope: A
  status report.
\newblock {\em Mathematical Social Sciences}, 23:67--80, DOI:
  \href{https://dx.doi.org/10.1016/0165-4896(92)90038-7}{\ttfamily
  10.1016/0165-4896(92)90038-7}.

\bibitem[Fukuda, 2004]{fukuda2004there}
Fukuda, K. (2004).
\newblock Frequently asked questions in polyhedral computation.
\newblock \url{http://www.cs.mcgill.ca/~fukuda/soft/polyfaq/polyfaq.html}.

\bibitem[Gelfand et~al., 1992]{gelfand1992bayesian}
Gelfand, A.~E., Smith, A. F.~M., and Lee, T.-M. (1992).
\newblock Bayesian analysis of constrained parameter and truncated data
  problems using {{Gibbs}} sampling.
\newblock {\em Journal of the American Statistical Association}, 87:523--532,
  DOI: \href{https://dx.doi.org/10.2307/2290286}{\ttfamily 10.2307/2290286}.

\bibitem[Ghosh, 2011]{ghosh2011objective}
Ghosh, M. (2011).
\newblock Objective priors: {{An}} introduction for frequentists.
\newblock {\em Statistical Science}, 26:187--202,
  \url{https://projecteuclid.org/euclid.ss/1312204006}.

\bibitem[Haaf and Rouder, in press]{haaf2019some}
Haaf, J.~M. and Rouder, J.~N. (in press).
\newblock Some do and some don{\textquoteright}t? {{Accounting}} for
  variability of individual difference structures.
\newblock {\em Psychonomic Bulletin \& Review}, DOI:
  \href{https://dx.doi.org/10.3758/s13423-018-1522-x}{\ttfamily
  10.3758/s13423-018-1522-x}.

\bibitem[Heck, in press]{heck2018caveat}
Heck, D.~W. (in press).
\newblock A caveat on the {{Savage}}-{{Dickey}} density ratio: {{The}} case of
  computing {{Bayes}} factors for regression parameters.
\newblock {\em British Journal of Mathematical and Statistical Psychology},
  DOI: \href{https://dx.doi.org/10.1111/bmsp.12150}{\ttfamily
  10.1111/bmsp.12150}.

\bibitem[Heck and Davis-Stober, 2019]{heck2019multinomineq}
Heck, D.~W. and Davis-Stober, C.~P. (2019).
\newblock {\em The {{R}} Package Multinomineq: {{Bayesian}} Inference for
  Inequality-Constrained Multinomial Models}.
\newblock \url{https://github.com/danheck/multinomineq}.

\bibitem[Heck et~al., 2017]{heck2017information}
Heck, D.~W., Hilbig, B.~E., and Moshagen, M. (2017).
\newblock From information processing to decisions: {{Formalizing}} and
  comparing probabilistic choice models.
\newblock {\em Cognitive Psychology}, 96:26--40, DOI:
  \href{https://dx.doi.org/10.1016/j.cogpsych.2017.05.003}{\ttfamily
  10.1016/j.cogpsych.2017.05.003}.

\bibitem[Heck et~al., in press]{heck2018quantifying}
Heck, D.~W., Overstall, A., Gronau, Q.~F., and Wagenmakers, E.-J. (in press).
\newblock Quantifying uncertainty in transdimensional {{Markov}} chain {{Monte
  Carlo}} using discrete {{Markov}} models.
\newblock {\em Statistics \& Computing}, DOI:
  \href{https://dx.doi.org/10.1007/s11222-018-9828-0}{\ttfamily
  10.1007/s11222-018-9828-0}.

\bibitem[Heck and Wagenmakers, 2016]{heck2016adjusted}
Heck, D.~W. and Wagenmakers, E.-J. (2016).
\newblock Adjusted priors for {{Bayes}} factors involving reparameterized order
  constraints.
\newblock {\em Journal of Mathematical Psychology}, 73:110--116, DOI:
  \href{https://dx.doi.org/10.1016/j.jmp.2016.05.004}{\ttfamily
  10.1016/j.jmp.2016.05.004}.

\bibitem[Heck et~al., 2015]{heck2015testing}
Heck, D.~W., Wagenmakers, E.-J., and Morey, R.~D. (2015).
\newblock Testing order constraints: {{Qualitative}} differences between
  {{Bayes}} factors and normalized maximum likelihood.
\newblock {\em Statistics \& Probability Letters}, 105:157--162, DOI:
  \href{https://dx.doi.org/10.1016/j.spl.2015.06.014}{\ttfamily
  10.1016/j.spl.2015.06.014}.

\bibitem[Hertwig et~al., 2004]{hertwig2004decisions}
Hertwig, R., Barron, G., Weber, E.~U., and Erev, I. (2004).
\newblock Decisions from experience and the effect of rare events in risky
  choice.
\newblock {\em Psychological Science}, 15:534--539, DOI:
  \href{https://dx.doi.org/10.1111/j.0956-7976.2004.00715.x}{\ttfamily
  10.1111/j.0956-7976.2004.00715.x}.

\bibitem[Hilbig and Moshagen, 2014]{hilbig2014generalized}
Hilbig, B.~E. and Moshagen, M. (2014).
\newblock Generalized outcome-based strategy classification: {{Comparing}}
  deterministic and probabilistic choice models.
\newblock {\em Psychonomic Bulletin \& Review}, 21:1431--1443, DOI:
  \href{https://dx.doi.org/10.3758/s13423-014-0643-0}{\ttfamily
  10.3758/s13423-014-0643-0}.

\bibitem[Hoijtink, 2011]{hoijtink2011informative}
Hoijtink, H. (2011).
\newblock {\em Informative Hypotheses: {{Theory}} and Practice for Behavioral
  and Social Scientists}.
\newblock {Chapman \& Hall/CRC}, {Boca Raton, FL}.

\bibitem[Hoijtink et~al., 2014]{hoijtink2014cognitive}
Hoijtink, H., B{\'{e}}land, S., and Vermeulen, J.~A. (2014).
\newblock Cognitive diagnostic assessment via {{Bayesian}} evaluation of
  informative diagnostic hypotheses.
\newblock {\em Psychological Methods}, 19:21--38, DOI:
  \href{https://dx.doi.org/10.1037/a0034176}{\ttfamily 10.1037/a0034176}.

\bibitem[Hoijtink et~al., 2008]{hoijtink2008bayesian}
Hoijtink, H., Klugkist, I., and Boelen, P.~A. (2008).
\newblock Bayesian evaluation of informative hypotheses.

\bibitem[Iverson and Falmagne, 1985]{iverson1985statistical}
Iverson, G. and Falmagne, J.-C. (1985).
\newblock Statistical issues in measurement.
\newblock {\em Mathematical Social Sciences}, 10:131--153, DOI:
  \href{https://dx.doi.org/10.1016/0165-4896(85)90031-9}{\ttfamily
  10.1016/0165-4896(85)90031-9}.

\bibitem[Iverson, 2006]{iverson2006essay}
Iverson, G.~J. (2006).
\newblock An essay on inequalities and order-restricted inference.
\newblock {\em Journal of Mathematical Psychology}, 50:215--219, DOI:
  \href{https://dx.doi.org/10.1016/j.jmp.2006.01.007}{\ttfamily
  10.1016/j.jmp.2006.01.007}.

\bibitem[Kahneman and Tversky, 1979]{kahneman1979prospect}
Kahneman, D. and Tversky, A. (1979).
\newblock Prospect theory: {{An}} analysis of decision under risk.
\newblock {\em Econometrica}, 47:263--291, DOI:
  \href{https://dx.doi.org/10.2307/1914185}{\ttfamily 10.2307/1914185}.

\bibitem[Karabatsos, 2001]{karabatsos2001rasch}
Karabatsos, G. (2001).
\newblock The {{Rasch}} model, additive conjoint measurement, and new models of
  probabilistic measurement theory.
\newblock {\em Journal of Applied Measurement}, 2:389--423.

\bibitem[Karabatsos, 2005]{karabatsos2005exchangeable}
Karabatsos, G. (2005).
\newblock The exchangeable multinomial model as an approach to testing
  deterministic axioms of choice and measurement.
\newblock {\em Journal of Mathematical Psychology}, 49:51--69, DOI:
  \href{https://dx.doi.org/10.1016/j.jmp.2004.11.001}{\ttfamily
  10.1016/j.jmp.2004.11.001}.

\bibitem[Karabatsos, 2018]{karabatsos2018bayesian}
Karabatsos, G. (2018).
\newblock On {{Bayesian}} testing of additive conjoint measurement axioms using
  synthetic likelihood.
\newblock {\em Psychometrika}, 83:321--332, DOI:
  \href{https://dx.doi.org/10.1007/s11336-017-9581-x}{\ttfamily
  10.1007/s11336-017-9581-x}.

\bibitem[Karabatsos and Sheu, 2004]{karabatsos2004orderconstrained}
Karabatsos, G. and Sheu, C.-F. (2004).
\newblock Order-constrained {{Bayes}} inference for dichotomous models of
  unidimensional nonparametric {{IRT}}.
\newblock {\em Applied Psychological Measurement}, 28:110--125, DOI:
  \href{https://dx.doi.org/10.1177/0146621603260678}{\ttfamily
  10.1177/0146621603260678}.

\bibitem[Klaassen et~al., 2018]{klaassen2018all}
Klaassen, F., Zedelius, C.~M., Veling, H., Aarts, H., and Hoijtink, H. (2018).
\newblock All for one or some for all? {{Evaluating}} informative hypotheses
  using multiple {{N}} = 1 studies.
\newblock {\em Behavior Research Methods}, 50:2276--2291, DOI:
  \href{https://dx.doi.org/10.3758/s13428-017-0992-5}{\ttfamily
  10.3758/s13428-017-0992-5}.

\bibitem[Klauer and Kellen, 2015]{klauer2015flexibility}
Klauer, K.~C. and Kellen, D. (2015).
\newblock The flexibility of models of recognition memory: {{The}} case of
  confidence ratings.
\newblock {\em Journal of Mathematical Psychology}, 67:8--25, DOI:
  \href{https://dx.doi.org/10.1016/j.jmp.2015.05.002}{\ttfamily
  10.1016/j.jmp.2015.05.002}.

\bibitem[Klauer et~al., 2015]{klauer2015parametric}
Klauer, K.~C., Singmann, H., and Kellen, D. (2015).
\newblock Parametric order constraints in multinomial processing tree models:
  {{An}} extension of {{Knapp}} and {{Batchelder}} (2004).
\newblock {\em Journal of Mathematical Psychology}, 64:1--7, DOI:
  \href{https://dx.doi.org/10.1016/j.jmp.2014.11.001}{\ttfamily
  10.1016/j.jmp.2014.11.001}.

\bibitem[Klugkist and Hoijtink, 2007]{klugkist2007bayes}
Klugkist, I. and Hoijtink, H. (2007).
\newblock The {{Bayes}} factor for inequality and about equality constrained
  models.
\newblock {\em Computational Statistics \& Data Analysis}, 51:6367--6379, DOI:
  \href{https://dx.doi.org/10.1016/j.csda.2007.01.024}{\ttfamily
  10.1016/j.csda.2007.01.024}.

\bibitem[Klugkist et~al., 2005a]{klugkist2005bayesian}
Klugkist, I., Kato, B., and Hoijtink, H. (2005a).
\newblock Bayesian model selection using encompassing priors.
\newblock {\em Statistica Neerlandica}, 59:57--69.

\bibitem[Klugkist et~al., 2005b]{klugkist2005inequality}
Klugkist, I., Laudy, O., and Hoijtink, H. (2005b).
\newblock Inequality constrained analysis of variance: {{A Bayesian}} approach.
\newblock {\em Psychological Methods}, 10:477.

\bibitem[Klugkist et~al., 2010]{klugkist2010bayesian}
Klugkist, I., Laudy, O., and Hoijtink, H. (2010).
\newblock Bayesian evaluation of inequality and equality constrained hypotheses
  for contingency tables.
\newblock {\em Psychological Methods}, 15:281--299, DOI:
  \href{https://dx.doi.org/10.1037/a0020137}{\ttfamily 10.1037/a0020137}.

\bibitem[Koppen, 1995]{koppen1995random}
Koppen, M. (1995).
\newblock Random utility representation of binary choice probabilities:
  {{Critical}} graphs yielding critical necessary conditions.
\newblock {\em Journal of Mathematical Psychology}, 39:21--39, DOI:
  \href{https://dx.doi.org/10.1006/jmps.1995.1003}{\ttfamily
  10.1006/jmps.1995.1003}.

\bibitem[Krantz et~al., 1971]{krantz1971foundations}
Krantz, D.~H., Luce, R.~D., Suppes, P., and Tversky, A. (1971).
\newblock {\em Foundations of Measurement - Volume 1}.
\newblock {Academic Press}, {New York, NY}.

\bibitem[Lange, 2010]{lange2010numerical}
Lange, K. (2010).
\newblock {\em Numerical Analysis for Statisticians}.
\newblock Statistics and {{Computing}}. {Springer}, {New York, NY}.

\bibitem[Laudy and Hoijtink, 2007]{laudy2007bayesian}
Laudy, O. and Hoijtink, H. (2007).
\newblock Bayesian methods for the analysis of inequality constrained
  contingency tables.
\newblock {\em Statistical Methods in Medical Research}, 16:123--138, DOI:
  \href{https://dx.doi.org/10.1177/0962280206071925}{\ttfamily
  10.1177/0962280206071925}.

\bibitem[Lee and Vanpaemel, 2018]{lee2018determining}
Lee, M.~D. and Vanpaemel, W. (2018).
\newblock Determining informative priors for cognitive models.
\newblock {\em Psychonomic Bulletin \& Review}, 25:114--127, DOI:
  \href{https://dx.doi.org/10.3758/s13423-017-1238-3}{\ttfamily
  10.3758/s13423-017-1238-3}.

\bibitem[Lindley, 1964]{lindley1964bayesian}
Lindley, D.~V. (1964).
\newblock The {{Bayesian}} analysis of contingency tables.
\newblock {\em The Annals of Mathematical Statistics}, 35:1622--1643,
  \url{https://www.jstor.org/stable/2238298}.

\bibitem[Lov{\'{a}}sz and Simonovits, 1993]{lovasz1993random}
Lov{\'{a}}sz, L. and Simonovits, M. (1993).
\newblock Random walks in a convex body and an improved volume algorithm.
\newblock {\em Random Structures \& Algorithms}, 4:359--412, DOI:
  \href{https://dx.doi.org/10.1002/rsa.3240040402}{\ttfamily
  10.1002/rsa.3240040402}.

\bibitem[Lov{\'{a}}sz and Vempala, 2006]{lovasz2006hitandrun}
Lov{\'{a}}sz, L. and Vempala, S. (2006).
\newblock Hit-and-run from a corner.
\newblock {\em SIAM Journal on Computing}, 35:985--1005, DOI:
  \href{https://dx.doi.org/10.1137/S009753970544727X}{\ttfamily
  10.1137/S009753970544727X}.

\bibitem[Marley and Regenwetter, 2017]{marley2017choice}
Marley, A. A.~J. and Regenwetter, M. (2017).
\newblock Choice, preference, and utility: {{Probabilistic}} and deterministic
  representations.
\newblock In Batchelder, W.~H., Colonius, H., Dzhafarov, E.~N., and Myung, J.,
  editors, {\em New {{Handbook}} of {{Mathematical Psychology}}}, volume~1,
  pages 374--453. {Cambridge University Press}, {Cambridge, MA}.

\bibitem[McCausland and Marley, 2013]{mccausland2013prior}
McCausland, W.~J. and Marley, A. A.~J. (2013).
\newblock Prior distributions for random choice structures.
\newblock {\em Journal of Mathematical Psychology}, 57:78--93, DOI:
  \href{https://dx.doi.org/10.1016/j.jmp.2013.05.001}{\ttfamily
  10.1016/j.jmp.2013.05.001}.

\bibitem[Meng, 1994]{meng1994posterior}
Meng, X.-L. (1994).
\newblock Posterior predictive p-values.
\newblock {\em The Annals of Statistics}, 22:1142--1160, DOI:
  \href{https://dx.doi.org/10.1214/aos/1176325622}{\ttfamily
  10.1214/aos/1176325622}.

\bibitem[Mulder et~al., 2012]{mulder2012biems}
Mulder, J., Hoijtink, H., and de~Leeuw, C. (2012).
\newblock {{BIEMS}}: {{A Fortran}} 90 program for calculating {{Bayes}} factors
  for inequality and equality constrained models.
\newblock {\em Journal of Statistical Software}, 46:1--39, DOI:
  \href{https://dx.doi.org/10.18637/jss.v046.i02}{\ttfamily
  10.18637/jss.v046.i02}.

\bibitem[Myung and Pitt, 1997]{myung1997applying}
Myung, I.~J. and Pitt, M.~A. (1997).
\newblock Applying {{Occam}}{\textquoteright}s razor in modeling cognition: {{A
  Bayesian}} approach.
\newblock {\em Psychonomic Bulletin \& Review}, 4:79--95, DOI:
  \href{https://dx.doi.org/10.3758/BF03210778}{\ttfamily 10.3758/BF03210778}.

\bibitem[Myung et~al., 2005]{myung2005bayesian}
Myung, J.~I., Karabatsos, G., and Iverson, G.~J. (2005).
\newblock A {{Bayesian}} approach to testing decision making axioms.
\newblock {\em Journal of Mathematical Psychology}, 49:205--225, DOI:
  \href{https://dx.doi.org/10.1016/j.jmp.2005.02.004}{\ttfamily
  10.1016/j.jmp.2005.02.004}.

\bibitem[Nunkesser et~al., 2009]{nunkesser2009rporta}
Nunkesser, R., Straatmann, S., Wenzel, S., Christof, T., and Loebel, A. (2009).
\newblock {\em {{rPorta}}: {{R}}/{{PORTA}} Interface}.
\newblock {R package version 0.1-93},
  \url{https://CRAN.R-project.org/package=rPorta}.

\bibitem[Paes et~al., 1997]{paes1997impact}
Paes, A. H.~P., Bakker, A., and Soe-Agnie, C.~J. (1997).
\newblock Impact of dosage frequency on patient compliance.
\newblock {\em Diabetes Care}, 20:1512--1517, DOI:
  \href{https://dx.doi.org/10.2337/diacare.20.10.1512}{\ttfamily
  10.2337/diacare.20.10.1512}.

\bibitem[Plummer, 2003]{plummer2003jags}
Plummer, M. (2003).
\newblock {{JAGS}}: {{A}} program for analysis of {{Bayesian}} graphical models
  using {{Gibbs}} sampling.
\newblock In {\em Proceedings of the 3\textsuperscript{rd} {{International
  Workshop}} on {{Distributed Statistical Computing}}}, volume 124, page 125.
  {Vienna, Austria}.

\bibitem[Prince et~al., 2012]{prince2012design}
Prince, M., Brown, S., and Heathcote, A. (2012).
\newblock The design and analysis of state-trace experiments.
\newblock {\em Psychological Methods}, 17:78--99, DOI:
  \href{https://dx.doi.org/10.1037/a0025809}{\ttfamily 10.1037/a0025809}.

\bibitem[Regenwetter and Cavagnaro, in press]{regenwetter2019tutorial}
Regenwetter, M. and Cavagnaro, D.~R. (in press).
\newblock Tutorial on removing the shackles of regression analysis: {{How}} to
  stay true to your theory of binary response probabilities.
\newblock {\em Psychological Methods}, DOI:
  \href{https://dx.doi.org/10.1037/met0000196}{\ttfamily 10.1037/met0000196}.

\bibitem[Regenwetter et~al., 2018]{regenwetter2018heterogeneity}
Regenwetter, M., Cavagnaro, D.~R., Popova, A., Guo, Y., Zwilling, C., Lim,
  S.~H., and Stevens, J.~R. (2018).
\newblock Heterogeneity and parsimony in intertemporal choice.
\newblock {\em Decision}, 5:63--94, DOI:
  \href{https://dx.doi.org/10.1037/dec0000069}{\ttfamily 10.1037/dec0000069}.

\bibitem[Regenwetter et~al., 2011]{regenwetter2011transitivity}
Regenwetter, M., Dana, J., and Davis-Stober, C.~P. (2011).
\newblock Transitivity of preferences.
\newblock {\em Psychological Review}, 118:42--56, DOI:
  \href{https://dx.doi.org/10.1037/a0021150}{\ttfamily 10.1037/a0021150}.

\bibitem[Regenwetter and Davis-Stober, 2012]{regenwetter2012behavioral}
Regenwetter, M. and Davis-Stober, C.~P. (2012).
\newblock Behavioral variability of choices versus structural inconsistency of
  preferences.
\newblock {\em Psychological Review}, 119:408--416, DOI:
  \href{https://dx.doi.org/10.1037/a0027372}{\ttfamily 10.1037/a0027372}.

\bibitem[Regenwetter and Davis-Stober, 2018]{regenwetter2018role}
Regenwetter, M. and Davis-Stober, C.~P. (2018).
\newblock The role of independence and stationarity in probabilistic models of
  binary choice.
\newblock {\em Journal of Behavioral Decision Making}, 31:100--114, DOI:
  \href{https://dx.doi.org/10.1002/bdm.2037}{\ttfamily 10.1002/bdm.2037}.

\bibitem[Regenwetter et~al., 2014]{regenwetter2014qtest}
Regenwetter, M., Davis-Stober, C.~P., Lim, S.~H., Guo, Y., Popova, A.,
  Zwilling, C., Cha, Y.-S., and Messner, W. (2014).
\newblock {{QTest}}: {{Quantitative}} testing of theories of binary choice.
\newblock {\em Decision}, 1:2--34, DOI:
  \href{https://dx.doi.org/10.1037/dec0000007}{\ttfamily 10.1037/dec0000007}.

\bibitem[Regenwetter and Robinson, 2017]{regenwetter2017constructbehavior}
Regenwetter, M. and Robinson, M.~M. (2017).
\newblock The construct--behavior gap in behavioral decision research: {{A}}
  challenge beyond replicability.
\newblock {\em Psychological Review}, 124:533--550, DOI:
  \href{https://dx.doi.org/10.1037/rev0000067}{\ttfamily 10.1037/rev0000067}.

\bibitem[Rissanen, 1978]{rissanen1978modeling}
Rissanen, J. (1978).
\newblock Modeling by shortest data description.
\newblock {\em Automatica}, 14:465--471, DOI:
  \href{https://dx.doi.org/10.1016/0005-1098(78)90005-5}{\ttfamily
  10.1016/0005-1098(78)90005-5}.

\bibitem[Robert and Casella, 2004]{robert2004monte}
Robert, C. and Casella, G. (2004).
\newblock {\em Monte {{Carlo}} Statistical Methods}.
\newblock {Springer Science \& Business Media}, {New York, NY}.

\bibitem[Robertson et~al., 1988]{robertson1988order}
Robertson, T., Wright, F.~T., and Dykstra, R.~L. (1988).
\newblock {\em Order Restricted Statistical Inference}.
\newblock {Wiley}, {Chichester, NY}.

\bibitem[Sanderson, 2010]{sanderson2010armadillo}
Sanderson, C. (2010).
\newblock Armadillo: {{An}} open source {{C}}++ linear algebra library for fast
  prototyping and computationally intensive experiments.

\bibitem[Sedransk et~al., 1985]{sedransk1985bayesian}
Sedransk, J., Monahan, J., and Chiu, H.~Y. (1985).
\newblock Bayesian estimation of finite population parameters in categorical
  data models incorporating order restrictions.
\newblock {\em Journal of the Royal Statistical Society. Series B
  (Methodological)}, 47:519--527, \url{https://www.jstor.org/stable/2345787}.

\bibitem[Silvapulle and Sen, 2004]{silvapulle2004constrained}
Silvapulle, M.~J. and Sen, P.~K. (2004).
\newblock {\em Constrained statistical inference: Order, inequality, and shape
  constraints}.
\newblock {Wiley}, {Hoboken, NJ}.

\bibitem[Smeulders et~al., 2018]{smeulders2018testing}
Smeulders, B., Davis-Stober, C., Regenwetter, M., and Spieksma, F. C.~R.
  (2018).
\newblock Testing probabilistic models of choice using column generation.
\newblock {\em Computers \& Operations Research}, 95:32--43, DOI:
  \href{https://dx.doi.org/10.1016/j.cor.2018.03.001}{\ttfamily
  10.1016/j.cor.2018.03.001}.

\bibitem[Smith, 1984]{smith1984efficient}
Smith, R.~L. (1984).
\newblock Efficient {{Monte Carlo Procedures}} for {{Generating Points
  Uniformly Distributed}} over {{Bounded Regions}}.
\newblock {\em Operations Research}, 32:1296--1308, DOI:
  \href{https://dx.doi.org/10.1287/opre.32.6.1296}{\ttfamily
  10.1287/opre.32.6.1296}.

\bibitem[{Stan Development Team}, 2018]{stan}
{Stan Development Team} (2018).
\newblock {\em Stan Modeling Language Users Guide and Reference Manual}.
\newblock {Version 2.18}, \url{http://mc-stan.org/}.

\bibitem[Stephan et~al., 2009]{stephan2009bayesian}
Stephan, K.~E., Penny, W.~D., Daunizeau, J., Moran, R.~J., and Friston, K.~J.
  (2009).
\newblock Bayesian model selection for group studies.
\newblock {\em NeuroImage}, 46:1004--1017, DOI:
  \href{https://dx.doi.org/10.1016/j.neuroimage.2009.03.025}{\ttfamily
  10.1016/j.neuroimage.2009.03.025}.

\bibitem[Suck, 1992]{suck1992geometric}
Suck, R. (1992).
\newblock Geometric and combinatorial properties of the polytope of binary
  choice probabilities.
\newblock {\em Mathematical Social Sciences}, 23:81--102, DOI:
  \href{https://dx.doi.org/10.1016/0165-4896(92)90039-8}{\ttfamily
  10.1016/0165-4896(92)90039-8}.

\bibitem[Wetzels et~al., 2010]{wetzels2010encompassing}
Wetzels, R., Grasman, R. P. P.~P., and Wagenmakers, E.-J. (2010).
\newblock An encompassing prior generalization of the {{Savage}}--{{Dickey}}
  density ratio.
\newblock {\em Computational Statistics \& Data Analysis}, 54:2094--2102, DOI:
  \href{https://dx.doi.org/10.1016/j.csda.2010.03.016}{\ttfamily
  10.1016/j.csda.2010.03.016}.

\end{thebibliography}
\end{document}